\documentclass[superscriptaddress,
 amsmath,amssymb,
 aps,
pra,
reprint
]{revtex4-2}

\usepackage{graphicx}
\usepackage{dcolumn}
\usepackage{bm}
\usepackage{physics}
\usepackage{amsfonts, amsmath}
\usepackage{dsfont}
\usepackage{amsthm}
\usepackage{hyperref}
\usepackage{bm}
\usepackage{bbold}
\usepackage{color}
\usepackage{lipsum,babel}
\usepackage[normalem]{ulem}
\usepackage{tabularx}

\newcommand{\haf}{\text{haf}}

\newcommand{\poly}{\text{poly}}

\begin{document}
\definecolor{navy}{RGB}{46,72,102}
\definecolor{pink}{RGB}{219,48,122}
\definecolor{grey}{RGB}{184,184,184}
\definecolor{yellow}{RGB}{255,192,0}
\definecolor{grey1}{RGB}{217,217,217}
\definecolor{grey2}{RGB}{166,166,166}
\definecolor{grey3}{RGB}{89,89,89}
\definecolor{red}{RGB}{255,0,0}

\preprint{APS/123-QED}

\title{Classical algorithm for simulating experimental Gaussian boson sampling}
\author{Changhun Oh}
\thanks{The authors contributed equally to this work.}
\affiliation{Pritzker School of Molecular Engineering, The University of Chicago, Chicago, IL 60637, USA}
\author{Minzhao Liu}
\thanks{The authors contributed equally to this work.}
\affiliation{Department of Physics, The University of Chicago, Chicago, IL 60637, USA}
\affiliation{Computational Science Division, Argonne National Laboratory, Lemont, IL 60439, USA}
\author{Yuri Alexeev}
\affiliation{Computational Science Division, Argonne National Laboratory, Lemont, IL 60439, USA}
\affiliation{Department of Computer Science, The University of Chicago, Chicago, IL 60637, USA}
\affiliation{Chicago Quantum Exchange, Chicago, IL 60637, USA}
\author{Bill Fefferman}
\affiliation{Department of Computer Science, The University of Chicago, Chicago, IL 60637, USA}
\author{Liang Jiang}
\affiliation{Pritzker School of Molecular Engineering, The University of Chicago, Chicago, IL 60637, USA}

\begin{abstract}
Gaussian boson sampling is a promising candidate for showing experimental quantum advantage.
While there is evidence that noiseless Gaussian boson sampling is hard to efficiently simulate using a classical computer, the current Gaussian boson sampling experiments inevitably suffer from loss and other noise models.
Despite a high photon loss rate and the presence of noise, they are currently claimed to be hard to classically simulate with the best-known classical algorithm.
In this work, we present a classical tensor-network algorithm that simulates Gaussian boson sampling and whose complexity can be significantly reduced when the photon loss rate is high.
By generalizing the existing thermal-state approximation algorithm of lossy Gaussian boson sampling, the proposed algorithm allows us to achieve increased accuracy as the running time of the algorithm scales, as opposed to the algorithm that samples from the thermal state, which can give only a fixed accuracy.
This generalization enables us to simulate the largest scale Gaussian boson sampling experiment so far using relatively modest computational resources, even though the output state of these experiments is not believed to be close to a thermal state.
By demonstrating that our new classical algorithm outperforms the large-scale experiments on the benchmarks used as evidence for quantum advantage, we exhibit evidence that our classical sampler can simulate the ground-truth distribution better than the experiment can, which disputes the experimental quantum advantage claims.
\end{abstract}

\maketitle

\section{Introduction}
We have seen the first plausible quantum computational advantage experiments over the past few years using random circuit sampling with superconducting qubits~\cite{arute2019quantum, wu2021strong, morvan2023phase} and Gaussian boson sampling~\cite{zhong2020quantum, zhong2021phase, madsen2022quantum, deng2023gaussian}. 
These experiments are not only an important step toward practical quantum advantage but also a fundamental milestone as evidence of violation of the extended Church--Turing thesis~\cite{bernstein1993quantum, aaronson2011computational}.
After many recent results (e.g., Refs.~\cite{aaronson2011computational, hamilton2017gaussian, boixo2018characterizing, bouland2019complexity, deshpande2021quantum}), we now have a basis for believing that these experiments are hard to classically simulate in the ideal case with no noise. 
While this is a necessary foundation, the actual experiments are very noisy, which could make the sampling problem a lot easier to classically simulate.
Hence, to understand the computational power of the existing experiments, we need to rigorously analyze the effect of noise on their complexity.
For this reason, numerous theoretical studies have been undertaken to understand the complexity of simulating noisy quantum devices~\cite{aharonov1996limitations, kalai2014gaussian, bremner2017achieving, gao2018efficient, oszmaniec2018classical, garcia2019simulating, noh2020efficient, qi2020regimes, oh2021classical, deshpande2022tight, aharonov2022polynomial, oh2023classical, hangleiter2023computational}.

More specifically, it is reasonable to expect that quantum systems with uncorrected noise become easy to simulate at a sufficiently large size, and this intuition has been justified by many recent algorithmic results (e.g., Refs.~\cite{aharonov1996limitations, kalai2014gaussian, aharonov2022polynomial, oh2023classical}).
Therefore, the uncorrected noise limits the scale of the quantum advantage experiments.
On the other hand, a sufficiently large-scale circuit is necessary to achieve  quantum advantage;
otherwise, the computational complexity of the problem that a quantum device solves is still tractable using a currently available classical computer.
Therefore, the hope of quantum advantage using noisy quantum circuits is to find the ``Goldilocks" regime where the system size is sufficiently large to be classically intractable but not too large for noise to destroy the quantum signal.
This requires understanding how well the best (possibly inefficient) classical algorithm can do to take advantage of noise.

\begin{figure*}[t]
\includegraphics[width=450px]{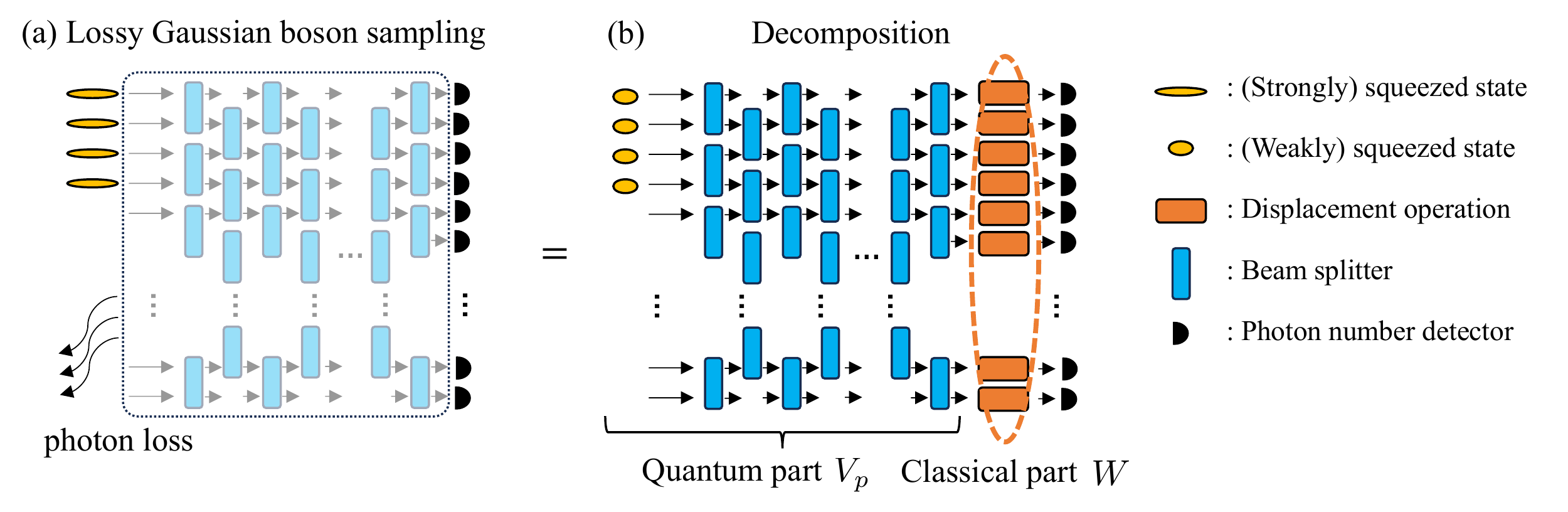}
\caption{(a) Gaussian boson sampling with input squeezed vacuum states that pass through a lossy beam splitter network. (b) Using the decomposition introduced in the main text, we decompose the output state as pure input squeezed states with reduced squeezing, followed by a lossless beam splitter network and Gaussian random displacement channel. Note that the random displacement follows a Gaussian distribution that is generally correlated over different modes.}
\label{fig:scheme}
\end{figure*}

In this work, we give a new classical tensor-network-based algorithm simulating Gaussian boson sampling that takes advantage of high-loss experiments which enables us to simulate current-size Gaussian boson sampling.
To do this, we revisit a family of algorithms that have previously appeared in the simulation literature and involve sampling from a classically easy thermal state~\cite{garcia2019simulating, qi2020regimes, martinez2023classical}, which do not currently work to simulate the existing experiments since the current Gaussian boson sampling experiments' output states are not believed to be close to a thermal state.
These algorithms sample from the distribution of thermal output state with covariance matrix ${V_\text{th}=W+\mathbb{1}_{2M}}$ for approximating the actual output state, where~$\mathbb{1}_{2M}$ represents a vacuum state and $W$ represents random displacement applied on the vacuum.
Our observation is that the output state of a Gaussian boson sampling experiment before measurement can always be written in a similar way, ${V=W+V_p}$, where~$W$ is again classical random displacement and $V_p$ is essentially equivalent to an ideal but smaller-photon-number Gaussian boson sampling's output covariance matrix~(see Fig.~\ref{fig:scheme}).
Based on this observation, our strategy is to exploit the tensor-network method to approximately simulate $V_p$, which enables us to approximate $V_p$ by $\mathbb{1}_{2M}$ in the trivial case, much like the thermal state approximation, and to increase the accuracy by increasing the bond dimension to converge to $V_p$.
The subsequent random displacement $W$ and measurement have greatly reduced the computational complexity for the tensor-network method because they can be operated locally.
Also, we observe that when a loss rate is high, the random displacement $W$ becomes dominant, and the number of photons in $V_p$ becomes very small.
Since the state $V_p$ has fewer photons, a tensor-network method can more efficiently simulate it than $V$, particularly when the loss rate is high. 

We employ the algorithm to analyze the computational power of the state-of-the-art Gaussian boson sampling experiments.
We simulate the ground-truth distribution of the largest Gaussian boson sampling experiments so far and verify that our algorithm outperforms the experiments for various benchmarks that were used to claim a quantum advantage in experiments.
(Throughout this work, the ground-truth means the ideal quantity obtained by the covariance matrix of the output state that includes photon loss, following the convention in the literature~\cite{villalonga2021efficient, madsen2022quantum, deng2023gaussian}.)
It provides evidence that our classical algorithm can simulate the ground-truth distribution better than the experimental samplers can.
Our classical simulation for the most time-consuming experiment takes about 10 minutes for constructing the matrix product state (MPS)
and about 62 minutes for sampling 10 million samples.
Therefore, based on benchmarking used in experiments, our tensor network algorithm can simulate the ground-truth distribution of Gaussian boson sampling better than the state-of-the-art experiments in a reasonable time.
We finally provide estimates of the computational cost of our algorithm for simulating larger systems by analyzing memory and time cost using the required bond dimension for MPS.

Our results provide a new, more accurate measure of the complexity of finite-size lossy Gaussian boson sampling.
A typical way of estimating the simulation cost for Gaussian boson sampling experiments is based on the best-known classical algorithm~\cite{bulmer2022boundary, quesada2022quadratic}, whose complexity substantially increases with the output photon number.
Because of this property, many previous experiments focused on increasing the output photon numbers.
However, our analysis clearly reveals that most of the output photons are classical, which do not contribute to the exponential cost, and shows that the quantum part $V_p$ is a more accurate quantum resource.
Hence, our algorithm indicates that a more promising path to the Goldilocks regime for quantum advantage is reducing the loss rate and increasing the number of squeezed states instead of focusing on increasing the output photon numbers.

The paper is structured as follows.
After providing a preliminary discussion of Gaussian boson sampling in Sec.~\ref{sec:pre}, we present our classical algorithm and the theory behind the algorithm in Sec.~\ref{sec:algorithm}.
We also analyze its scaling as the system size increases and the simulation accuracy.
In Sec.~\ref{sec:numerical}, we provide simulation results of the recent Gaussian boson sampling experiments.
In Sec.~\ref{sec:estimate}, we provide estimates of computational cost for larger Gaussian boson sampling experiments for future experiments.
In Sec.~\ref{sec:discussion}, we discuss the implications of our results and open questions.


\section{Gaussian boson sampling}\label{sec:pre}
Gaussian boson sampling is a sampling task that is proven to be hard to classically simulate for noiseless circuits under some plausible conjectures~\cite{aaronson2011computational, hamilton2017gaussian, deshpande2021quantum}.
The standard Gaussian boson sampling can be implemented by first preparing squeezed vacuum states and injecting them into an $M$-mode beam-splitter circuit, characterized by an ${M\times M}$ unitary matrix $U$ (see Fig.~\ref{fig:scheme}~(a) without loss effect.).
Then the output state's photon numbers at each mode are measured.
The crucial observation for hardness is that computing the output probability of Gaussian boson sampling is \#P-hard, which is written as
\begin{align}
    p(\bm{m})=\frac{1}{\prod_{i=1}^M \cosh r_i}\frac{|\haf (A_{\bm{m}})|^2}{m_1!\cdots m_M!},
\end{align}
where $\bm{m}=(m_1,\dots,m_M)$ is an output photon number pattern, $A$ is defined as $A=UDU^\text{T}$,  $D$ is a diagonal matrix defined as $D=\text{diag}(\{r_i\}_{i=1}^M)$ with squeezing parameters $r_i$'s, $A_{\bm{m}}$ is the matrix obtained by repeating the $i$th row and columns of the matrix $A$ $m_i$ times, and the hafnian of an $n\times n$ matrix $X$ is defined as
\begin{align}
    \haf(X)=\sum_{m\in \text{PMP}(n)}\prod_{(i,j)\in m}X_{i,j}
\end{align}
where PMP($n$) represents the set of perfect matching permutations of $n$ (even) elements.

\section{Classical simulation algorithm}\label{sec:algorithm}

\subsection{Decomposition of the output state of Gaussian boson sampling}\label{sec:decomposition}
We now present a decomposition of the Gaussian boson sampling's output state, which is a crucial first step for our classical algorithm.
As mentioned, our strategy is to decompose the output state into the quantum and classical parts.
To do that, we decompose the output Gaussian state's covariance matrix into two parts as $V=V_p+W$, where $V_p$ represents the covariance matrix of a pure Gaussian state and $W\succeq 0$, which is illustrated in Fig.~\ref{fig:scheme} (see Sec.~\ref{methods:cov} for details of covariance matrix formalism).
Here, the covariance matrix $V_p$ can be interpreted as a pure quantum resource because it is composed of pure squeezed states and beam splitters, which is equivalent to the standard Gaussian boson sampling with much smaller squeezing parameters than the input of $V$.
On the other hand, the positive semidefinite matrix $W$ can be interpreted as a Gaussian random displacement because the initial covariance matrix $V$ can be obtained by applying Gaussian random displacement characterized by the classical covariance matrix $W$ to the pure Gaussian state of its covariance matrix $V_p$~\cite{serafini2017quantum}.

As emphasized before, one may interpret our classical algorithm as a generalization of the thermal-state approximation algorithm~\cite{garcia2019simulating, qi2020regimes} because if we approximate the quantum part $V_p$ as a vacuum state~$\mathbb{1}_{2M}$, then the corresponding simulation becomes a thermal-state approximation with covariance matrix~${W+\mathbb{1}_{2M}}$.
Such a generalization allows us to improve the approximate simulation error by increasing the running time (see Sec.~\ref{sec:mps}), which is crucial for simulating current experiments, the output state of which is not sufficiently close to a thermal state.
Although Refs.~\cite{qi2020regimes, martinez2023classical} employ a classical {\it squeezed} thermal state, we call this state a thermal state for simplicity throughout the paper.

\begin{figure}[t]
\includegraphics[width=240px]{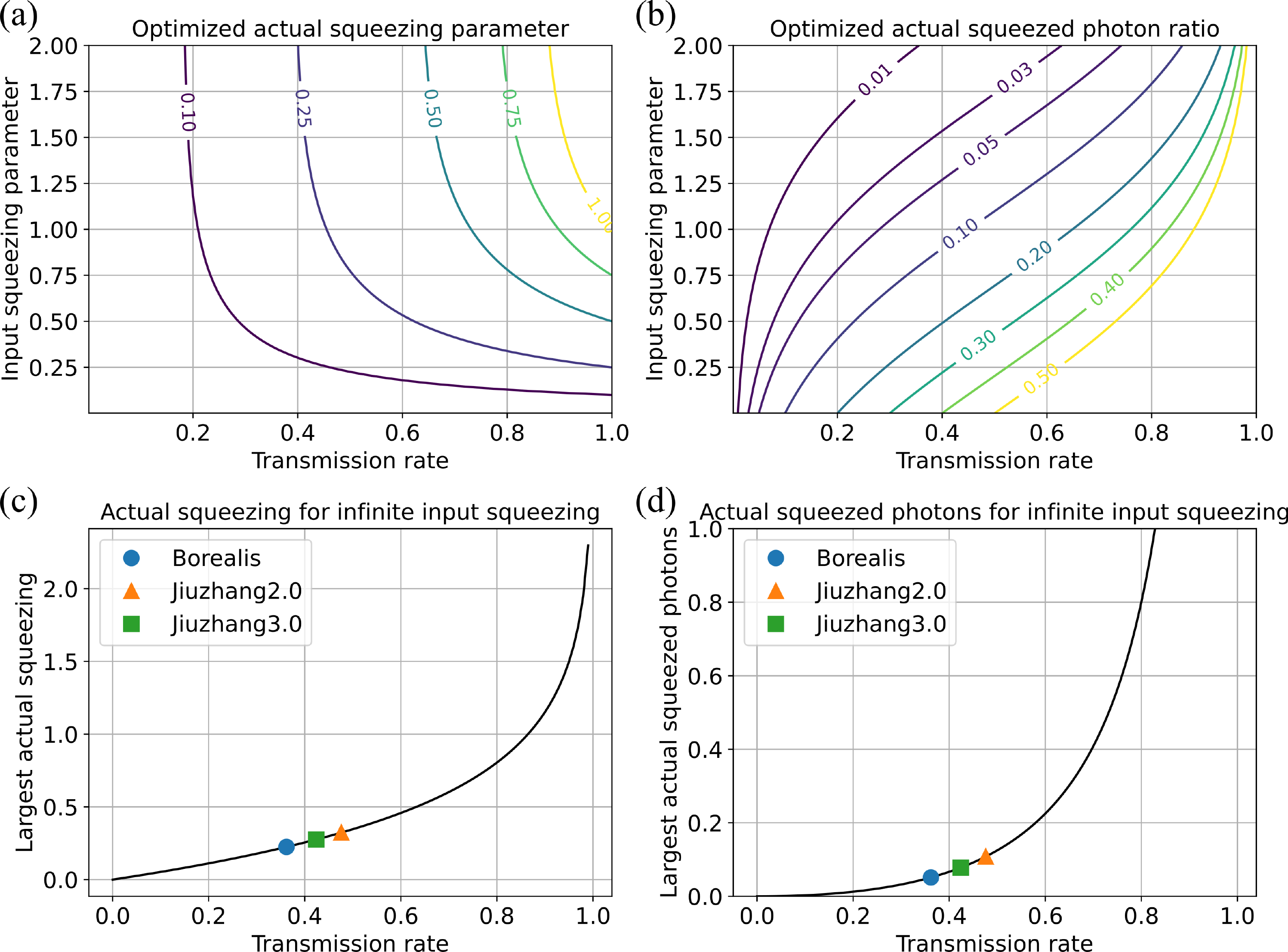} 
\caption{Characteristics of the squeezed state $V_p$ from the decomposition for single-mode cases. (a) Relation between the actual squeezing parameter $s$ and the input squeezing $r$ for different transmission rate $\eta$. (b) Ratio of the actual squeezed photons to the total photons of the output state $V$. (c) and (d) Actual squeezing parameter and squeezed photon numbers when the input squeezing parameter is infinite.
The dots represent the Borealis, Jiuzhang2.0, and Jiuzhang3.0's circuit's transmission rate and their largest actual squeezing and squeezed photons, assuming that infinite input squeezing is used.}
\label{fig:decompose}
\end{figure}

\begin{table*}
\begin{ruledtabular}
\begin{tabular}{p{0.85in} p{0.6in} p{0.5in} p{0.3in} p{0.3in} p{0.63in} p{0.3in} p{0.45in} p{0.35in}}
    Dataset & Experiment & Input squeezing & Input photons & Trans. rate & Actual squeezing & Output photons & Actual squeezed photons & Ratio \\
    \hline
    \hline
    M16 & Borealis & 0.88-0.89 & 16.248 & 0.368 & 0.14-0.22 & 5.98 & 0.549 & 0.0917\\
    \hline
    M72 & Borealis & 0.88-0.89 & 72.77 & 0.317 & 0.093-0.2 & 23.056 & 1.74 & 0.0755\\
    \hline
    M216 (low) & Borealis & 0.52-0.54 & 67.38 & 0.321 & 0.06-0.154 & 21.622 & 3.09 & 0.143\\
    \hline
    M216 (high) & Borealis & 1.09-1.11 & 388.57 & 0.324 & 0.087-0.235 & {\bf 125.85} & {\bf 6.54} & 0.052 \\ 
    \hline
    M288 & Borealis & 1.00-1.02 & 407.64 & 0.362 & 0.102-0.247 & {\bf 147.65} & {\bf 10.687} & 0.0724
    \\ 
    \hline
    M100 & Jiuzhang1.0 & 1.35-1.84 & 277.64 & 0.283 & 0.08-0.26 & 78.62 & 1.5 & 0.019 \\ 
    \hline
    M144 (P125-1) & Jiuzhang2.0 & 0.47-0.56 & 14.593 & 0.539 & 0.16-0.255 & 7.87 & 2.337 & 0.297  \\ 
    \hline
    M144 (P125-2) & Jiuzhang2.0 & 0.72-0.94 & 43.09 & 0.538 & 0.215-0.342 & 23.189 & 4.281 & 0.1846 \\
    \hline
    M144 (P65-1) & Jiuzhang2.0 & 0.4-0.545 & 13.415 & 0.476 & 0.124-0.217 & 6.39 & 1.628 & 0.255  \\
    \hline
    M144 (P65-2) & Jiuzhang2.0 & 0.56-0.76 & 28.267 & 0.476 & 0.154-0.270 & 13.454 & 2.53 & 0.188 \\
    \hline
    M144 (P65-3) & Jiuzhang2.0 & 0.80-1.08 & 65.86 & 0.476 & 0.18-0.32 & 31.34 & 3.62 & 0.115 \\
    \hline
    M144 (P65-4) & Jiuzhang2.0 & 1.04-1.41 & 133.75 & 0.476 & 0.20-0.36 & 63.636 & 4.385 & 0.069 \\
    \hline
    M144 (P65-5) & Jiuzhang2.0 & 1.34-1.81 & 295.15 & 0.476 & 0.212-0.379 & {\bf 140.38} & {\bf 4.965} & 0.035 \\
    \hline
    M144 (low) & Jiuzhang3.0 & 1.14-1.26 & 113.04 & 0.424 & 0.185-0.299 & 47.93 & 3.08 & 0.064 \\
    \hline
    M144 (median) & Jiuzhang3.0 & 1.33-1.47 & 183.46 & 0.424 & 0.193-0.314 & 77.80 & 3.37 & 0.043 \\
    \hline
    M144 (high) & Jiuzhang3.0 & 1.49-1.66 & 274.22 & 0.424 & 0.198-0.323 & {\bf 116.29} & {\bf 3.556} & 0.031
\end{tabular}
\end{ruledtabular}
\caption{Parameters of different Gaussian boson sampling experiments from Refs.~\cite{zhong2020quantum, zhong2021phase, madsen2022quantum, deng2023gaussian}. We display the actual squeezing parameters and the actual squeezed photons for each experiment obtained by the optimal decomposition introduced in the main text.}
\label{table:parameters}
\end{table*}

More specifically, for the multimode Gaussian state's covariance matrix $V$, we can implement the optimized decomposition procedure by using semidefinite programming under the constraints
\begin{align}
    \min_{V_p} \Tr[V_p]~\text{with}~V-V_p\succeq0,~ V_p\succeq i\Omega,
\end{align}
where the first constraint is to ensure that $W=V-V_p$ is positive semidefinite, and the second constraint is to guarantee that $V_p$ represents a proper physical Gaussian state's covariance matrix, that is,  corresponds to a positive semidefinite density matrix~\cite{serafini2017quantum}.
Note that this is different from a simple Williamson decomposition used in Ref.~\cite{quesada2022quadratic} (see Appendix~\ref{app:single}).
Here, the minimization is to minimize the quantum part's average photon number.

We apply the method to the recent Gaussian boson sampling experiments' ground-truth covariance matrices~\cite{zhong2020quantum, zhong2021phase, madsen2022quantum}, which are given in Table~\ref{table:parameters}.
Here the actual squeezed photons represent the mean photon numbers from the covariance matrix $V_p$, namely, $\Tr[V_p-\mathbb{1}_{2M}]/4$.
One can clearly see that the resultant squeezed photons are much smaller than the total number of photons.
For example, for the largest experiment in Ref.~\cite{madsen2022quantum} with $M=288$, although the total output photon number was $147.65$, the true quantum resources are only around $7$ percent, that is, around $11$ photons.
Another interesting feature is that in Ref.~\cite{zhong2021phase}, although Jiuzhang2.0's different experiments used different squeezing parameters, which led to a significant difference for output photon numbers, the actual squeezed photons did not substantially change much because the transmission rate was fixed.
This highlights the importance of improving the transmission rate to increase quantum resources.
Since the low transmission rate highly limits the actual squeezing parameters in experiments, increasing the number of squeezed states is a better way to increase the actual squeezed photon number.
Indeed, whereas Jiuzhang2.0's actual squeezed photons are not significantly increased by increasing input squeezing parameters, Borealis's actual squeezed photons grow faster by increasing the number of squeezed states.


To understand this better, we quantitatively analyze the effect of loss for the single-mode case.
For different input squeezing parameters $r\geq 0$ and transmission rates $\eta$, we analyze the resultant (smaller) squeezing parameter $s\geq 0$ of $V_p$ from the decomposition, which is exhibited in Fig.~\ref{fig:decompose}.
Here, the decomposition of a lossy squeezed state's covariance matrix $V$ is written as (see Appendix~\ref{app:single} for more details)
\begin{align}
    V
    &=
    \eta V_0+(1-\eta)\mathbb{1}_2 \\ 
    &=
    \begin{pmatrix}
        e^{2s} & 0 \\
        0 & e^{-2s}
    \end{pmatrix}
    +
    \begin{pmatrix}
        \eta e^{2r}+1-\eta-e^{2s} & 0 \\
        0 & 0
    \end{pmatrix}
    \equiv V_p+W,
\end{align}
where $V_0$ is the covariance matrix of the input before loss and $e^{-2s}\equiv \eta e^{-2r}+(1-\eta)$.
One can clearly see that when the loss rate is high, increasing the input squeezing parameter does not increase the resultant squeezing parameter $s$ and only increases thermal photons from the random displacement part $W$.
One can easily show that for a fixed transmission rate $\eta$, even if the input squeezing is infinite, that is, $r\to \infty$, the actual squeezing is only $s=-1/2 \log(1-\eta)$, which is shown in Figs.~\ref{fig:decompose}(c)(d).
Therefore, increasing the squeezing parameter without improving the transmission rate does not increase the quantum resources.

In summary, the proposed decomposition separates the output photons into two contributions: the quantum resource from $V_p$ and the classical resource from the random displacement $W$.
Intuitively, only the quantum resource can significantly increase the simulation complexity.
However, the currently available classical algorithm's complexity still significantly increases with the classical resource from the random displacement part (see Appendix~\ref{app:comparison} for detailed comparison with existing classical algorithms).
We now present a classical algorithm that exploits the decomposition, whose complexity does not increase with the random displacement as we desire.

\subsection{Matrix product state}\label{sec:mps}
The idea of exploiting the decomposition introduced in the preceding section is to simulate the quantum part using MPS.
An MPS is a useful tensor network tool that is frequently used to simulate a many-body quantum state~\cite{cirac2021matrix}.
In general, MPS enables us to describe a quantum state efficiently if its entanglement is not large~\cite{vidal2003efficient}.
As we have observed in the preceding section, when a loss rate is high, the resultant squeezing parameters are highly suppressed, which are necessary to constitute entanglement; thus, MPS can be expected to describe the system efficiently.
An MPS of a given pure quantum state is written as follows:
\begin{align}
|\psi\rangle&=\sum_{n_1,\cdots,n_M=0}^{d-1}c_{n_1\cdots n_M}|n_1,\cdots, n_M\rangle  \\ 
&\approx\sum_{n_1,\cdots,n_M=0}^{d-1}\sum_{\alpha_1,\cdots,\alpha_{M-1}=0}^{\chi-1}\Gamma_{\alpha_1}^{[1]n_1}\lambda_{\alpha_1}^{[1]}\Gamma_{\alpha_1\alpha_2}^{[2]n_2}\lambda_{\alpha_2}^{[2]}\times \nonumber \\ 
&~~~~~~~~\cdots \lambda_{\alpha_{M-1}}^{[M-1]}\Gamma_{\alpha_{M-1}}^{[M]n_M}|n_1,\cdots, n_M\rangle,
\end{align}
where $d$ is the dimension of a local Hilbert space and $\chi$ is the bond dimension.

\begin{figure*}[t!]
\includegraphics[width=420px]{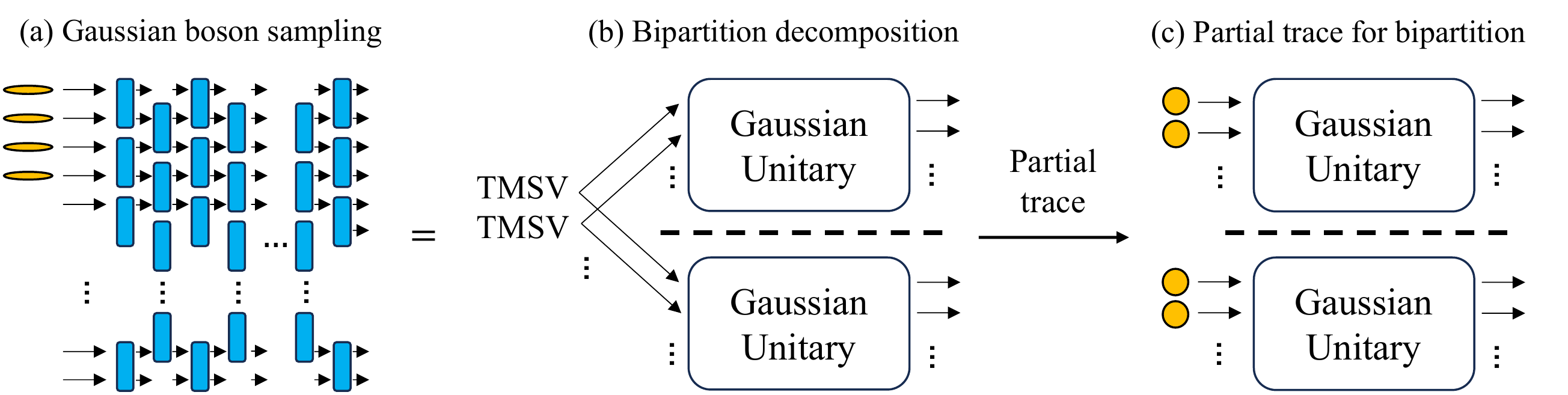}
\caption{(a) Any pure output state of a Gaussian boson sampling circuit can be decomposed as (b) the product of two-mode squeezed vacuum (TMSV) states followed by local Gaussian unitary operations. (c) Thus, after tracing out the other system, each local system can be described by the product of thermal states followed by Gaussian unitary operation. The Gaussian unitary operations for each system are different in general.}
\label{fig:bipartition}
\end{figure*}

A typical way to construct the output state's MPS description is to employ the time evolution method~\cite{oh2021classical, liu2023simulating}.
In this work, we present a method that directly constructs an MPS of a Gaussian state following the method from Ref.~\cite{vidal2003efficient}, which can significantly reduce the cost and the error occurring during time evolution.
While the method from this reference is typically inefficient in practice, thanks to the property of Gaussian states, the entire procedure takes only $O(cMd\chi^2)$, where $c$ is a parameter that depends on the system's characteristics, which will be described below.
We provide the details in Appendix~\ref{app:MPS} and summarize the result here:
\begin{align}
    &\Gamma^{[1]n_1}_{\alpha_1}=\langle n_1^{[1]}|\langle \bm{n}_{\alpha_1}^{[2\cdots M]}|\hat{U}^{[2\cdots M]\dagger}\hat{U}^{[1\cdots M]}|\bm{n}=\bm{0}\rangle/\lambda^{[1]}_{\alpha_1},\\
    &\Gamma^{[k]n_k}_{\alpha_{k-1}\alpha_{k}} \nonumber\\
    &=\langle n_k^{[k]}|\langle \bm{n}_{\alpha_k}^{[(k+1)\cdots M]}|\hat{U}^{[(k+1)\cdots M]\dagger}\hat{U}^{[k\cdots M]}|\bm{n}_{\alpha_{k-1}}^{[k\cdots M]}\rangle/\lambda^{[k]}_{\alpha_k} \nonumber \\ &~~~~~~~~~~~~~~~~~~~~~~~~~~~~~~~~~~~~~\text{for}~~ 1<k<M, \\
    &\Gamma^{[M]n_M}_{\alpha_{M-1}}=\langle n_M^{[M]}|\hat{U}^{[M]}|n_{\alpha_{M-1}}^{[M]}\rangle/\lambda^{[M-1]}_{\alpha_{M-1}},
\end{align}
where $|\bm{n}^{[k\cdots M]}\rangle$ represents a photon-number state over $[k\cdots M]$ modes, $\hat{U}^{[k\cdots M]}$ represents the Gaussian unitary operation diagonalizing the reduced density matrix of $[k\cdots M]$ modes, and each matrix element and singular values can be obtained by Williamson decomposition~\cite{serafini2017quantum}.
As shown in Appendix~\ref{app:MPS}, computing the matrix elements requires computing a hafnian of a matrix whose size is determined by $n_k^{[k]}+\sum_{i}(\bm{n}_{\alpha_k}^{[(k+1)\cdots M]})_i+\sum_{i}(\bm{n}_{\alpha_{k-1}}^{[k\cdots M]})_i$~\cite{quesada2019franck, oh2022quantum}, which corresponds to the aforementioned parameter~$c$.
Thus the presented MPS construction method's complexity increases exponentially in the latter (see Appendix~\ref{app:MPS} for more details).
Nonetheless, because of the high reduction of photon numbers from $V$ to $V_p$, the matrix size of which we compute the hafnian is substantially smaller than directly simulating $V_p$ using the best-known classical algorithms~\cite{bulmer2022boundary, quesada2022quadratic}.

Once we construct an MPS for the output state $V_p$ of the beam splitter circuit, the subsequent random displacement $W$ and photon number measurement can be implemented locally.
Hence, it does not increase entanglement anymore, and the sampling procedure can be efficiently conducted.
This part is the main difference from other algorithms~\cite{bulmer2022boundary, quesada2022quadratic} because our algorithm's complexity does not increase by the thermal photons or displacement.
Therefore, the complexity originates mainly from the quantum resources and is determined by the quantum part's entanglement because the remaining part is local.
In practice, in order to deal with the infinite dimensionality of continuous-variable systems, the local Hilbert space dimension $d$ has to be appropriately chosen.
Nonetheless, since the Gaussian states' photon number population has an exponential tail~\cite{hamilton2017gaussian}, truncating the local Hilbert space dimension entails only an exponentially small error.

More specifically, to simulate the random displacement channel, after sampling a random displacement from the covariance matrix $W$, we apply a random displacement on each site and sample from the displaced output state.
The idea is that thanks to the decomposition $V=V_p+W$, the ground-truth probability distribution can be written as
\begin{align}
    p(\bm{m})
    =\int d^{2M}\bm{\beta}p_W(\bm{\beta})p(\bm{m}|\bm{\beta}),
\end{align}
where $p_W(\bm{\beta})$ is the Gaussian probability distribution of the covariance matrix $W$ and $\beta$ is its random variable~${\beta\in\mathbb{C}^M}$ corresponding to a random displacement.
Therefore, sampling~$\bm{m}$ from the probability distribution~$p(\bm{m})$ is equivalent to sampling~$(\bm{m})$ from $p(\bm{m}|\bm{\beta})$ with randomly generated $\bm{\beta}$ from $p_W(\bm{\beta})$.
To sample from $p(\bm{m}|\bm{\beta})$ for a given random displacement $\bm{\beta}$, we transform the output MPS tensors by applying displacements $\bm{\beta}$ as
\begin{align}
    \sum_{n_k=0}^{d-1}\Gamma_{\alpha_{k-1}\alpha_k}^{[k-1]n_k}|n_k\rangle
    &\to \sum_{n_k=0}^{d-1}\hat{D}(\beta_k)\Gamma_{\alpha_{k-1}\alpha_k}^{[k-1]n_k}|n_k\rangle \\ 
    &=\sum_{n_k=0}^{d-1}\sum_{m_k=0}^{d-1}\Gamma_{\alpha_{k-1}\alpha_k}^{[k-1]n_k}\langle m_k|\hat{D}(\beta_k)|n_k\rangle |m_k\rangle \\ 
    &\equiv \sum_{m_k=0}^{d-1}\tilde{\Gamma}_{\alpha_{k-1}\alpha_k}^{[k-1]m_k}|m_k\rangle,
\end{align}
for each $1\leq k\leq M$, where $\hat{D}(\beta_k)$ is the displacement operator on the $k$th mode by $\beta_k$.
After transformation, we use the chain rule of marginal probabilities for a given random displacement $\bm{\beta}$ for sampling:
\begin{align}
    &p(m_1,\dots,m_M|\bm{\beta}) \nonumber \\ 
    &~~~~~~=p(m_1|\bm{\beta})\frac{p(m_1,m_2|\bm{\beta})}{p(m_1|\bm{\beta})}\cdots \frac{p(m_1,\dots,m_M|\bm{\beta})}{p(m_1,\dots,m_{M-1}|\bm{\beta})},
\end{align}
which completes the sampling procedure.


\subsection{Asymptotic behavior of MPS for lossy Gaussian boson sampling}\label{sec:entropy}
We now investigate when an efficient MPS representation exists, i.e., the required bond dimension is $\chi=\poly(K,1/\epsilon)$, where $\epsilon$ is an approximate error set by the bond dimension~and $K$ is the system size.
To this end, it suffices to investigate the R{\'e}nyi entropy of its reduced density matrix obtained by bipartition (see Sec.~\ref{methods:mps})~\cite{verstraete2006matrix, schuch2008entropy}.
We focus on $K$ two-mode squeezed input states distributed to bipartite systems because any multimode Gaussian states can be decomposed into a product of two-mode squeezed states up to bipartition~\cite{botero2003modewise}, as illustrated in Fig.~\ref{fig:bipartition}, and because the maximum entropy is achieved when each part of two-mode squeezed vacuum states is allocated to distinct bipartition.
We set squeezing parameters to be equal to $r\geq 0$ for all $K$ two-mode squeezed states for simplicity.

With the decomposition from Sec.~\ref{sec:decomposition}, it suffices to study the output state $V_p$ before random displacement in Fig.~\ref{fig:scheme}, because the rest of the operations are local.
Here, we are interested in the asymptotic regime with a loss rate converging to one as the system size scales, i.e., $\eta\to 0$, ($\eta=O(K^{-\beta})$ with $\beta>0$).
For this case, the squeezing parameter of $V_p$ becomes 
\begin{align}
    s=-\frac{1}{2}\log\left(\eta e^{-2r}+1-\eta\right)
    \approx \eta e^{-r}\sinh{r}+O(\eta^2),
\end{align}
where we fix $r$ to be a constant, and we can show in the asymptotic limit in $K$ (see Sec.~\ref{methods:mps}) that the R{\'e}nyi entropy of its reduced density matrix obtained by bipartition is given by
\begin{align}
    KS_{\alpha<1}(s)&=O(K\eta^{2\alpha}), \\ 
    KS_{\alpha=1.01}(s)&=\Omega(K\eta^{2}).
\end{align}
Hence, when $\eta=O((\log K/K)^{1/2\alpha})$ with $\alpha<1$, the MPS is efficient; that is, the required bond dimension scales as $\poly(K,1/\epsilon)$.
Note that when $\alpha\to 1$, the scaling approaches to $\eta=o((\log K/K)^{1/2})$, which recovers previous results $\eta=o(1/\sqrt{K})$~\cite{oszmaniec2018classical, garcia2019simulating, qi2020regimes} with an additional $(\log K)^{1/2}$ factor improvement.
Besides the logarithmic factor, a more important improvement is that our classical algorithm can control the approximation error $\epsilon$ in $\poly(1/\epsilon)$ by increasing the bond dimension $\chi$, whereas  the thermal state approximation cannot control the error by increasing the running time.
In addition, we show in Appendix~\ref{app:error} that for any fixed circuits, the required bond dimension to achieve an error $\epsilon$ scales as $\chi=O(\text{polylog}(1/\epsilon))$.

On the other hand, $KS_{\alpha=1.01}(s)=\Omega(K\eta^{2})=\Omega(K^\gamma)$ with any constant $\gamma>0$ when $\eta=\Omega(K^{(\gamma-1)/2})$, implying that the MPS algorithm starts to be inefficient~\cite{verstraete2006matrix}.
When $\eta=\Theta(1/\sqrt{K})$, we show that while a certain constant level of approximation error is attainable, the error may not be reduced efficiently (see Appendix~\ref{app:error}).

\section{Numerical implementation for finite-size circuits}\label{sec:numerical}
In the preceding section we analyzed our algorithm's asymptotic behavior.
In this section we investigate the efficiency of our algorithm for the state-of-the-art Gaussian boson sampling experiments~\cite{zhong2020quantum, zhong2021phase, madsen2022quantum, deng2023gaussian}.
We provide evidence that our algorithm can simulate the ground-truth distribution better than the existing experiments can  in a reasonable time.

\subsection{Benchmarking}
Before presenting our numerical results, let us briefly introduce the  benchmarking methods widely used for Gaussian boson sampling and how we verify our simulation results.
A standard theoretical metric is the TVD between the ground-truth distribution and the distribution of an experiment or a classical algorithm because it is the basis of the hardness evidence~\cite{aaronson2011computational, hamilton2017gaussian, deshpande2021quantum} and it operationally quantifies the difficulty of discriminating two samplers.
TVD cannot be estimated in practice, however, because it is neither sample-efficient nor computationally efficient.
Therefore, an indirect method invented to assess the TVD is XEB~\cite{boixo2018characterizing}, which is designed to quantify the weight of samples that have a large ideal probability (see Sec.~\ref{methods:XEB}). XEB was initially introduced and has been extensively studied for random circuit sampling with qubits~\cite{gao2021limitations, boixo2018characterizing, arute2019quantum, aaronson2016complexity, aaronson2019classical, morvan2023phase, ware2023sharp}.
While XEB is still computationally inefficient because we need to compute the ideal output probability, it is a sample-efficient measure, which is the reason that it was used in recent experiments~\cite{zhong2020quantum, madsen2022quantum}.
Another widely used benchmarking method in Gaussian boson sampling is the two-point~\cite{phillips2019benchmarking} or higher-order correlation function~\cite{zhong2021phase, deng2023gaussian}, which is defined recursively as
\begin{align}
    \kappa(n_1,n_2,\dots,n_k)
    =\mathbb{E}[n_1n_2\dots n_k]-\sum_{p\in P_k}\prod_{b\in p}\kappa[(n_i)_{i\in b}]
\end{align}
with $\kappa(n_i)\equiv \mathbb{E}[n_i]$ for a single element, where $P_k$ represents all the partitions of $\{1,2,\dots,k\}$, except the universal set, and the average is over the samples or the ground-truth distribution.
The purpose of correlation functions is to compare a given sampler's correlations with the ground-truth values and  analyze the similarity.

While XEB and correlation functions are widely employed for demonstrating the quantum advantage in experiments, neither method is proven to be rigorous for demonstrating quantum advantage, to the best of our knowledge.
In particular, a spoofing algorithm can achieve a large XEB for current experiments without simulating the ground-truth distribution~\cite{oh2023spoofing}.
 Vallalonga~et~al.~\cite{villalonga2021efficient} presented a classical sampler reproducing ideal two-point correlation functions without reproducing higher-order correlations.
Hence, it motivated experimentalists to further analyze high-order correlations to provide additional evidence than that of two-point or low-order correlations~\cite{zhong2021phase, deng2023gaussian}.
Nonetheless, it is still unclear whether the experimental samplers that lose some two-point correlations but reproduce higher correlations better have a smaller TVD than the algorithm in Ref.~\cite{villalonga2021efficient}.~(see Ref.~\cite{aaronson2021} for a related discussion)

Although we still employ the benchmarking methods despite their apparent drawbacks, we emphasize that our algorithm is not a spoofing algorithm but genuinely an approximate algorithm; that is, it is not designed to spoof benchmarking scores, unlike ones in Refs.~\cite{villalonga2021efficient, oh2023spoofing}.
In addition, we exploit those methods because those benchmarking methods are the basis of experimental quantum advantage claims.
Also, by analyzing small-size and intermediate-size experiments, we numerically observe an agreement between the TVD and XEB and between the XEB and two-point correlation functions.
Based on this observation, we analyze two-point correlation functions for the largest Gaussian boson sampling experiments and show that our algorithm provides better two-point correlation functions than the experiments provide.
We verify that our algorithm can also reproduce higher-order correlation functions, unlike the algorithm in Ref.~\cite{villalonga2021efficient}.
Therefore, we show that our simulator outperforms for the benchmarking methods that were evidence of quantum advantage.

\begin{figure}[t!]
\includegraphics[width=240px]{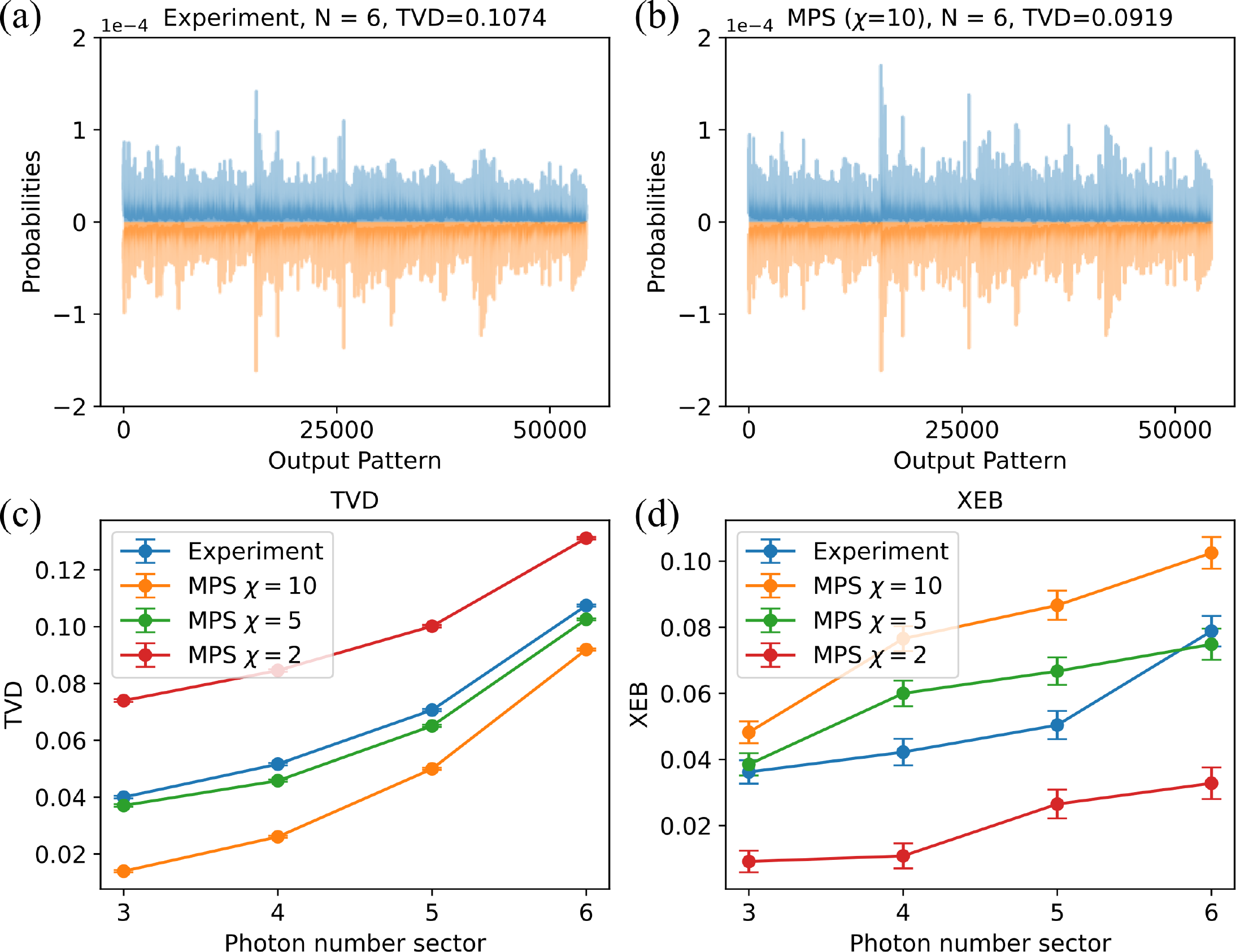}
\caption{(a)(b) Example output probability distributions. (c) TVD and (d) XEB for different photon number sectors. Here, for the TVD we used the empirically obtained probability distribution with 1 million samples for each sector, and we used 10,000 samples for XEB for each sector.
They clearly show the agreement between the XEB and TVD.
The error bar is obtained by 1,000 bootstrapping resamples.}
\label{fig:TVD_vs_XEB}
\end{figure}

\subsection{Small size}
We first simulate the small-size experiment from Ref.~\cite{madsen2022quantum}.
Since the experiment's Hilbert space dimension is small, we can compute all the probabilities and the TVD between the probability distributions obtained by samples and the ground-truth distribution.
We implement the MPS simulation with different bond dimensions $\chi=2,5,10$, which lead to different MPS truncation errors, $0.099, 0.033, 0.008$, respectively, and show the results in Fig.~\ref{fig:TVD_vs_XEB}.
For $\chi=5$, although we have lost around $3$ percent of squared singular values, the simulation accuracy is comparable to that of the experiment.
This is due to additional experimental noises, such as partial distinguishability, which makes the experimental sampler's output probability distribution deviate from the ground-truth distribution.
Therefore, it clearly shows that the experiments suffer from noises other than photon loss and lose the sampling accuracy to the ground-truth distribution, suggesting that the corresponding classical sampler does not need to achieve a very high precision to surpass the current experiments.

We now study the XEB and the relation between the TVD and XEB to employ the XEB for larger systems as a proxy of TVD, where we cannot efficiently obtain TVD.
Figures~\ref{fig:TVD_vs_XEB}(c)(d) clearly exhibit that the XEB and TVD follow the same tendency: that is, when the MPS's TVD is larger than the experiment, its XEB is smaller than the experiment, and vice versa.
Using this observation for our cases, we will use the XEB as a proxy of TVD for intermediate scales.

\begin{table*}
\begin{ruledtabular}
\begin{tabular}{p{0.56in} p{0.41in} p{1.3in} p{1.4in} p{1.3in} p{1.3in}}
    Dataset & Bond $\chi$ & Slope (Exp./MPS) & Correlation (Exp./MPS) & Distance (Exp./MPS) & Truncation error\\
    \hline
    \hline
    B-M72 & 120 & {\bf0.877}/0.861 & 0.977/{\bf0.984} & 0.049/0.049 & 0.048\\
    \hline
    B-M72 & 160 & 0.877/{\bf0.884} & 0.977/{\bf0.989} & 0.049/\bf{0.043} & 0.040\\
    \hline
    B-M72 & 200 & 0.877/{\bf0.899} & 0.977/{\bf0.990} & 0.049/\bf{0.039} & 0.032\\
    \hline
    B-M72 & 240 & 0.877/{\bf0.907} & 0.977/{\bf0.991} & 0.049/\bf{0.036} & 0.026\\
    \hline    
    B-M216-l & 600 & 0.919/{\bf0.935} & 0.936/\bf{0.952} & 0.021/\bf{0.018} & 0.012\\
    \hline
    B-M216-h & 10000 & 0.935/{\bf 0.972} & 0.964/\bf{0.980} & 0.199/\bf{0.151} & 0.006\\ 
    \hline
    B-M288 & 10000 & 0.887/\bf{0.937} & 0.960/\bf{0.970} & 0.207/\bf{0.197} & 0.017\\ 
    \hline
    J2-P65-1 & 1000 & 0.943/0.943 & 0.977/\bf{0.991} & 0.007/\bf{0.005} & 0.008\\
    \hline
    J2-P65-2 & 2000 & 0.936/\bf{0.939} & 0.981/\bf{0.993} & 0.015/\bf{0.010} & 0.017\\
    \hline
    J2-P65-3 & 10000 & 0.927/\bf{0.968} & 0.986/\bf{0.996} & 0.030/\bf{0.017} & 0.014\\
    \hline
    J2-P65-4 & 10000 & 0.927/\bf{0.972} & 0.988/\bf{0.997} & 0.048/\bf{0.025} & 0.025\\
    \hline
    J2-P65-5 & 10000 & 0.902/\bf{0.980} & 0.989/\bf{0.998} & 0.067/\bf{0.029} & 0.036\\
    \hline
    J3-high & 10000 & 0.954/\bf{0.982} & 0.993/\bf{0.998} & 0.048/\bf{0.026} & 0.014
\end{tabular}
\end{ruledtabular}
\caption{Two-point correlation function benchmarking for different scales of experiments. We present the slope, the correlation, and the two-norm distance to the ground-truth distribution's two-point correlations. We highlight the better scores.}
\label{table:twopoint}
\end{table*}

We note that in Refs.~\cite{zhong2021phase, madsen2022quantum, deng2023gaussian}, the Bayesian test has been used as another benchmarking method together with the XEB and correlation functions.
In this work we do not employ the Bayesian test because the test's purpose is essentially to compare  the likelihood of the ground-truth distribution with a mock-up distribution for a given experimental data; in other words, the score measures a distance from an experimental sampler to a mock-up distribution or the ideal sampler, not a distance from the ground-truth distribution to an experimental sampler or a mock-up sampler.
We provide more discussion in Appendix~\ref{app:bayesian}.

\begin{figure}[t!]
\includegraphics[width=230px]{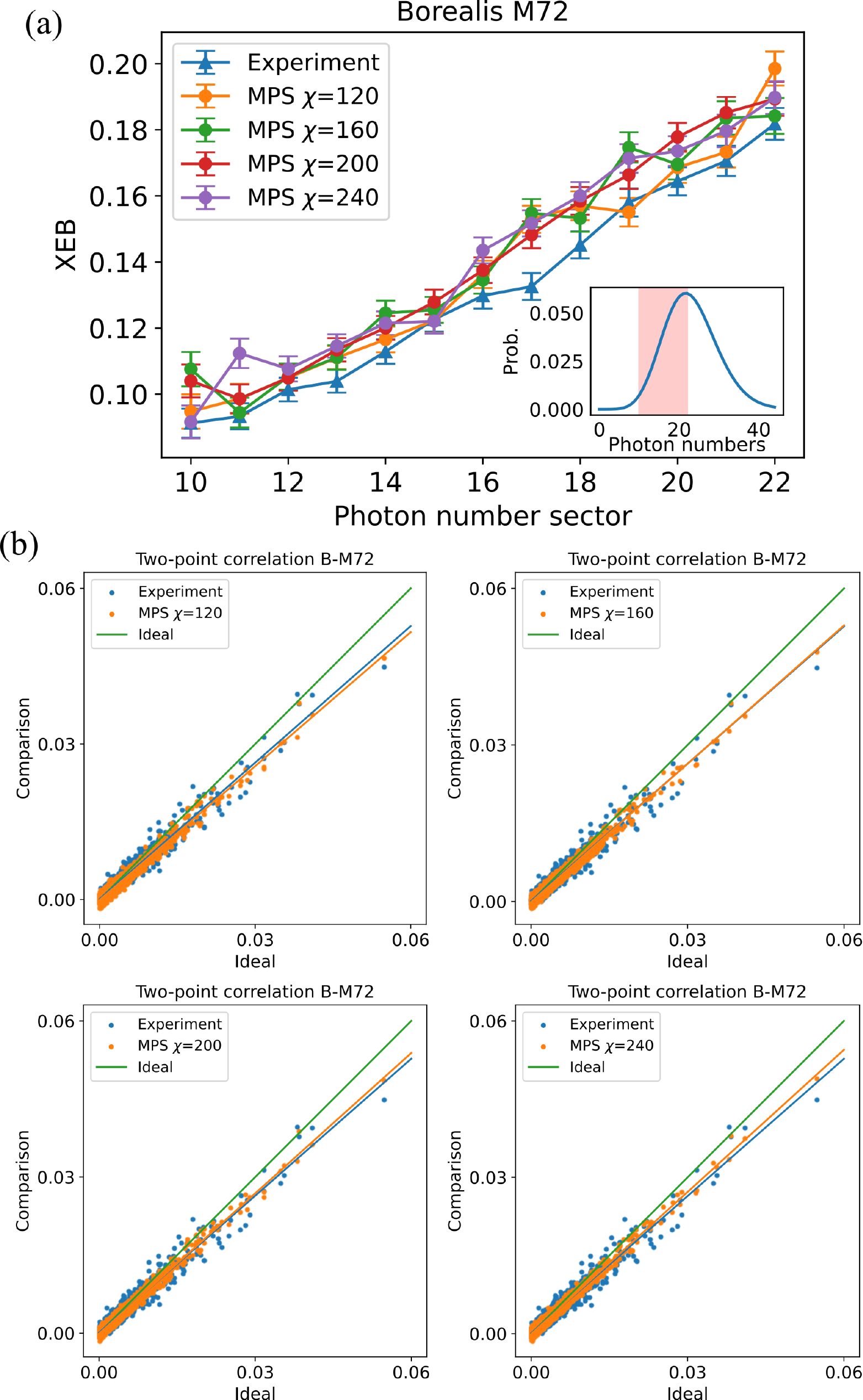} 
\caption{Simulation results of Borealis $M=72$ case with the MPS algorithm. (a) XEB;  (b) two-point correlation with different bond dimensions $\chi=120,160,200,240$. For the two-point correlation function calculation, we have used 1 million samples for all cases. The inset of (a) represents the total photon number distribution, and the shaded region is the sectors we used for XEB.
The error bar is obtained by 1,000 bootstrapping resamples.}
\label{fig:M72}
\end{figure}

\subsection{Intermediate size}
Now we implement the MPS algorithm for intermediate-size experiments, which were used to extrapolate for verifying the quantum advantage of their largest experiments against various mock-up distributions because results of intermediate-size experiments can still be verifiable using reasonable computational resources.
Unlike the previous small-size case, we can no longer compute the TVD because the number of outcomes is too large.
Therefore, based on the observation that the XEB may be a proxy of TVD, we will focus on the XEB.

We focus on Borealis's intermediate-scale experiment with $M=72$ and provide similar simulation results for different intermediate-scale experiments in Appendix~\ref{app:intermediate}.
We choose the bond dimensions as $\chi=120,160,200,240$ with local dimension $d=6$, which render the truncation error $0.048, 0.039, 0.032, 0.026$, respectively.
After sampling, we implement XEB for different photon number sectors, as shown in Fig.~\ref{fig:M72}(a).
One can clearly see that, overall, the bond dimensions we chose render larger XEB scores than the experiments do.
We then analyze the two-point correlation functions of all pairs of 72 modes and compare them with the ground-truth values,  presented in Fig.~\ref{fig:M72}(b).
We can see  that as the bond dimension increases, the two-point correlation functions are closer to the ideal cases.
Also, from $\chi=160$, the linear fit of the correlation functions with respect to the ideal case has a better slope than the experimental correlation functions have, which indicates that from $\chi=160$ our MPS sampler achieves better scores in the two-point correlation function benchmarking method in terms of the slope as well as XEB.
We also analyze other statistical quantities such as Pearson correlation of two-point correlations and two-norm distance of correlations between a sampler and the ground-truth's.
As shown in Table~\ref{table:twopoint}, from $\chi=160$, the MPS algorithm achieves a larger correlation and smaller distance.
The additional quantities may explain the reason that the XEB of the MPS with $\chi=120$ is better than the experiment, even though the former has a smaller slope in the two-point correlation's linear fit.
Note that although we could not compute the XEBs for all photon number sectors because of exponential cost for larger photon number sectors, the photon number sectors we analyze are typically the dominant photon numbers in general.

\begin{figure}[t]
\includegraphics[width=240px]{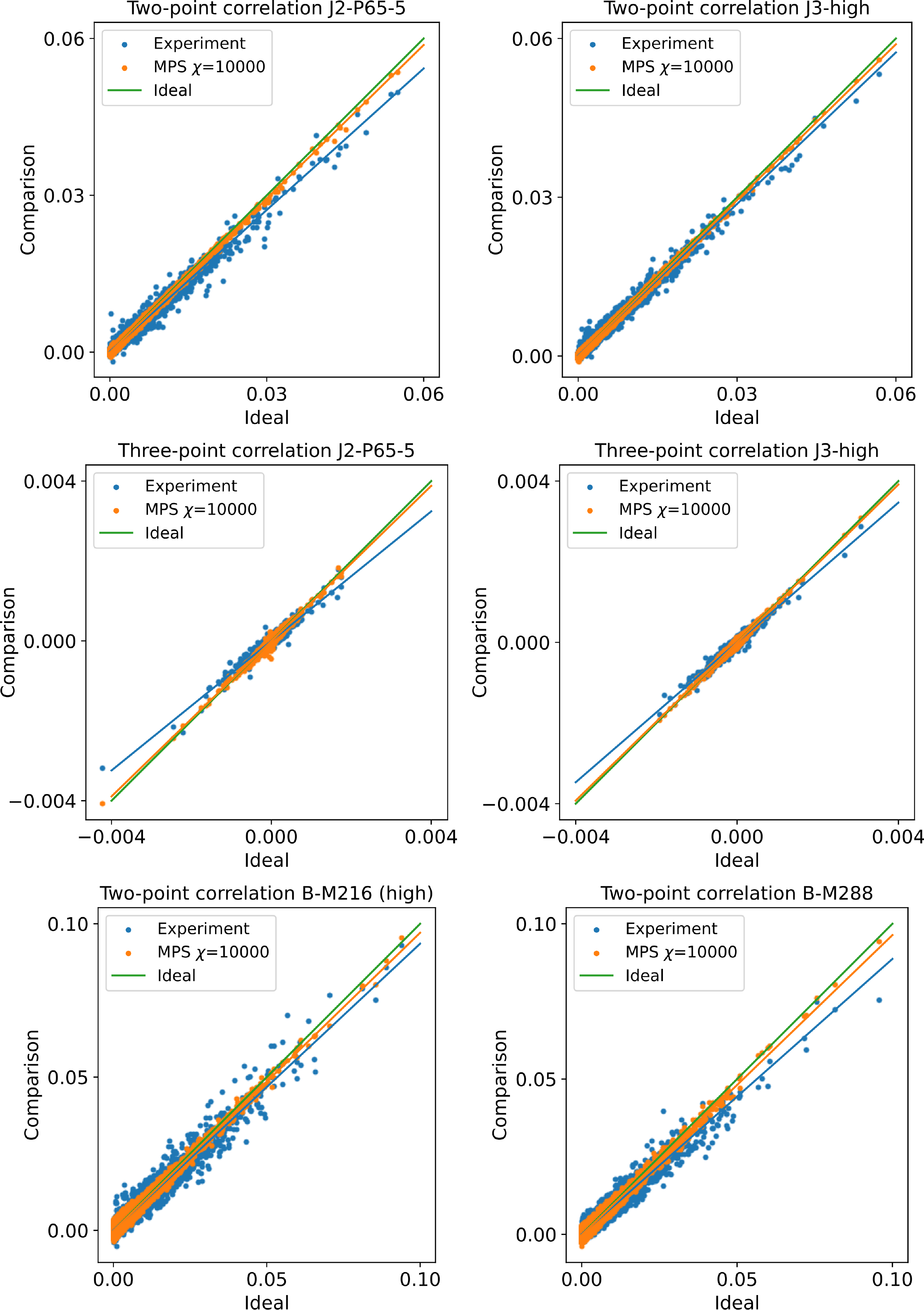} 
\caption{XEB and two-point correlation functions of experiments and our MPS sampler for Jiuzhang2.0's P65-5 with $M=144$, Jiuzhang3.0's high with $M=144$, and Borealis $M=216$ (high) and $M=288$. For two-point correlation functions, we use 1 million samples, and for three-point correlation functions we use 10 million samples.}
\label{fig:large}
\end{figure}

\begin{figure}[t]
\includegraphics[width=210px]{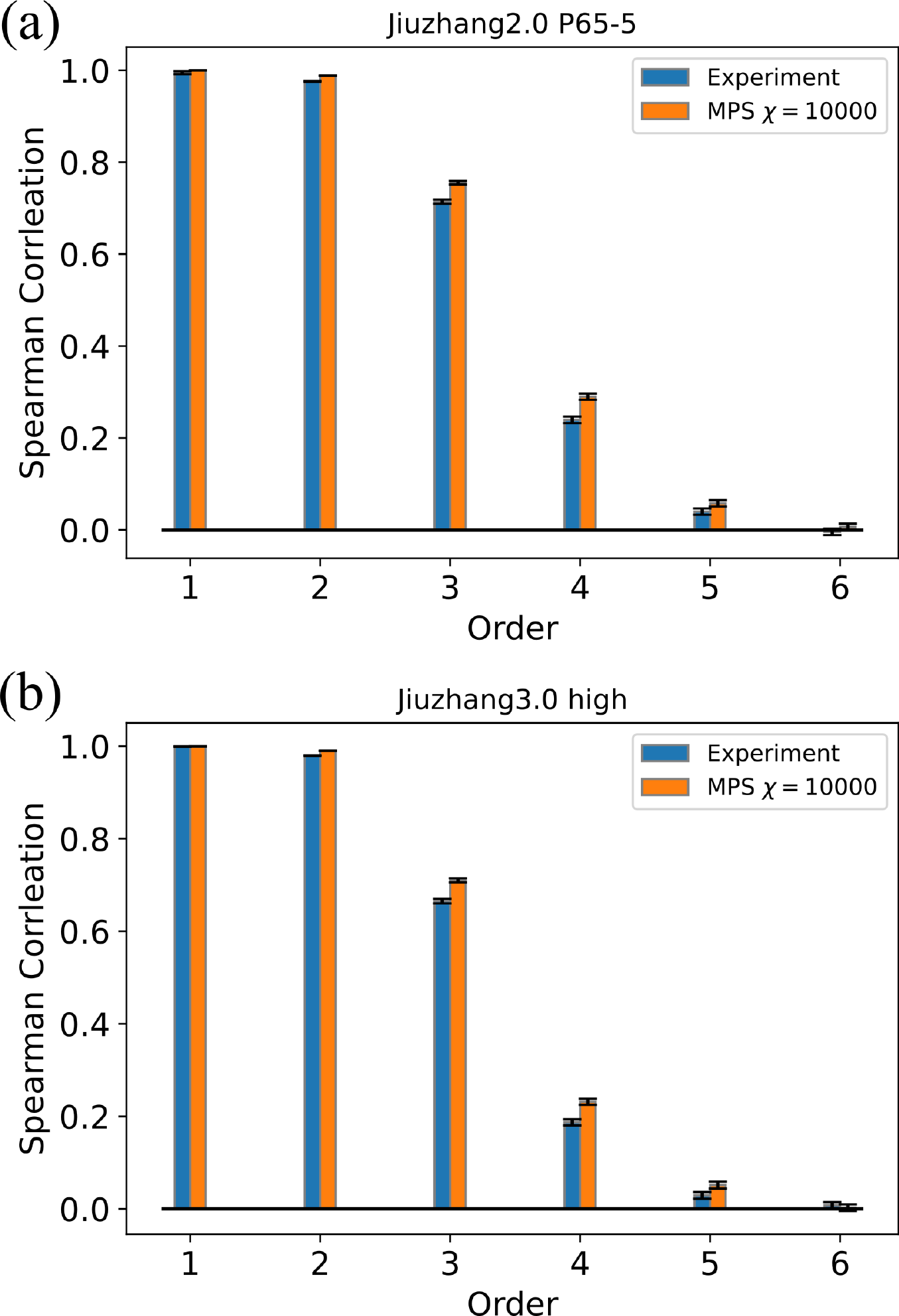}
\caption{Spearman correlation of samples' higher-order correlations to the ground-truth correlations. We use 20 million  samples for both samplers; and for each order, up to 20,000 randomly chosen subsets of modes out of $M=144$ modes were considered. For the first and second orders, we used all subsets. The error bars are the standard deviation obtained by 1,000 bootstrapping resamples.}
\label{fig:higher}
\end{figure}

\subsection{Largest scale}
We now simulate the largest Borealis, Jiuzhang2.0, and Jiuzhang3.0 experiments, which were used to claim quantum computational advantage.
For the benchmark we use two-point correlation functions because of the computationally large cost for XEB and the fact that they follow a similar tendency.
Here we choose the bond dimension $\chi=10000$ for all the cases and the cutoff $d=4$ for the MPS construction and $d=10$ for sampling.
Figure~\ref{fig:large} clearly shows that our classical algorithm performs significantly better than the experiments in terms of the slope of the linear fit.
Furthermore, one can see that our sampler's correlation functions have much smaller fluctuations for all the cases. In fact, if we look at the other metrics in Table~\ref{table:twopoint}, such as Pearson correlation of two-point correlations or two-norm distance to ground-truth values, our sampler consistently achieves much better scores.

We further analyze the higher-order correlation functions for Jiuzhang2.0 and Jiuzhang3.0, which are used as other benchmarks in Refs.~\cite{zhong2021phase, deng2023gaussian}.
As two-point correlation functions, we plot the third-order correlation functions in Fig.~\ref{fig:large}.
Clearly, the samples from MPS have a stronger correlation to the ground-truth correlation functions.
We further analyze the higher-order correlations up to the 6th order, as illustrated in Fig.~\ref{fig:higher}.
Here we present the Spearman correlation instead of Pearson correlation for consistency with Refs.~\cite{zhong2021phase, deng2023gaussian} while we observed a similar trend for Pearson correlations.
Up to the 5th order, the MPS samples' correlations manifestly correlate more with the ground-truth values, which is different from the spoofing algorithm from Ref.~\cite{villalonga2021efficient} for low-order correlation functions.
For the 6th order, although Jiuzhang3.0's case has a slightly larger correlation than the MPS samples have, the difference still lies within the error bar.
Therefore, up to the 6th order, we did not observe a clear advantage from experiments over our classical simulator.
We did not conduct the same analysis of higher-order correlations for Borealis experiments because the number of provided samples in Ref.~\cite{madsen2022quantum} is insufficient for higher-order correlation analysis.

Note that Jiuzhang3.0 applied 8-fold local beam splitters to each output mode for implementing pseudo-photon-number-resolving detection (PPNRD), which makes $M=144$ output modes into $M=1152$ modes.
In our simulation we treat the experiment as $M=144$ modes instead of $M=1152$ because the purpose of additional linear-optical circuits is to implement PPNRD and they are local beam splitters.
In other words, after generating samples by photon numbers, we transform the photon numbers to photon clicks for both the experiment and our sampler to compute the correlation functions of photon clicks for 144 modes.
We consider threshold detectors instead of PNRDs because the photon number for each output port is not dilute enough to mimic the true PNRD in experiment, especially for the high case.
We remark that Ref.~\cite{deng2023gaussian} modeled the ground-truth distribution as a lossy and partial distinguishable Gaussian boson sampler, additionally incorporating the partial distinguishability noise into the ground truth, while we still employ the lossy Gaussian boson sampler as the ground truth.
Hence, for our analysis, the partial distinguishability causes a deviation between the experimental and ground-truth samplers, which allows the MPS to have a larger truncation error.
We expect that if we include additional noises into the ground-truth distribution, the truncation error of MPS has to decrease further; that is, the bond dimension has to be chosen larger than the present simulation because the ground-truth distribution becomes closer to the experimental sampler.
However, the partial distinguishable Gaussian boson sampling model used in Ref.~\cite{deng2023gaussian} can be easily implemented by constructing two independent MPSs for two distinguishable input sources.
In addition, since a single-mode squeezed state is split into two smaller squeezed states, the cost of each MPS is smaller than the indistinguishable Gaussian boson sampler model.

\begin{figure}[t]
\includegraphics[width=200px]{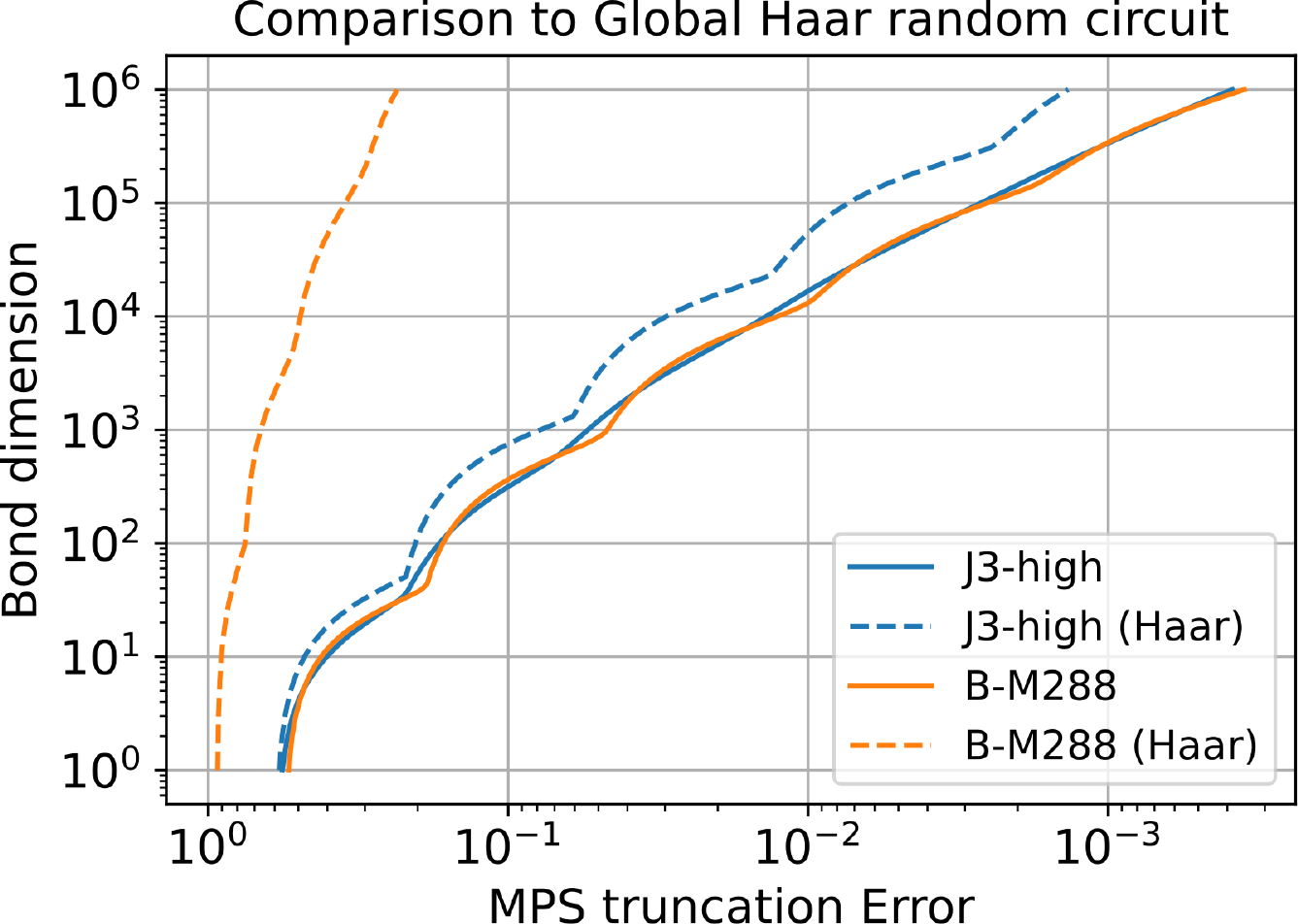} 
\caption{Comparison of required bond dimensions from the implemented experiments' circuits (solid curves) and when a global Haar-random circuit is implemented (dashed curves).}
\label{fig:error}
\end{figure}

We also analyze the required bond dimension for different truncation errors, shown in Fig.~\ref{fig:error}.
The truncation error in this figure is defined as the sum of the squared lost singular values for the bipartition at the center for simplicity.
Since the trace distance between the true state and the approximate MPS extensively depends on the lost singular values for other bipartitions~\cite{verstraete2006matrix, schuch2008entropy}, the error may not be consistent for different sizes of experiments (especially when the number of modes is different for each); thus, we do not compare different experiments.
The most important implication from the figure is that the MPS truncation error is another crucial parameter for its running time, which is related to experimental noises causing discrepancy from the ground-truth distribution because it allows the MPS to have truncations error to outperform the experiment.
It highlights that besides the experimental scales, the level of experimental noise significantly changes the cost of classical simulation.
We again emphasize that for a fixed-size circuit, the bond dimension scales as $\chi=O(\text{polylog}(1/\epsilon))$ in the TVD $\epsilon$, as shown in Sec.~\ref{sec:entropy}.
Another interesting feature is that assuming that each experiment implemented a global Haar-random linear-optical circuit, the required bond dimension substantially changes for Borealis but it does not for Jiuzhang.
This indicates that the connectivity of Jiuzhang is sufficiently large to implement an approximate global Haar-random circuit, whereas Borealis still requires a much deeper circuit.
From a simulation perspective, it also highlights that unlike the exact Gaussian boson sampling classical algorithm~\cite{quesada2022quadratic, bulmer2022boundary}, our classical algorithm inherently takes advantage of limited connectivity and the amount entanglement~\cite{vidal2003efficient}, as do the algorithms proposed in Refs.~\cite{oh2022classical, qi2022efficient}.

To simulate the largest experiments, we use the Polaris supercomputer at the Argonne Leadership Computing Facility. Each node has a single 2.8 GHz AMD EPYC Milan 7543P 32 core CPU, 512 GB of DDR4 RAM, four NVIDIA A100 GPUs connected via NVLink, and a pair of local 1.6TB of SSDs. The same number of GPUs as the number of optical modes (e.g., 288 GPUs for B-M288) are used in all simulations. The primary limitation on the bond dimension is the GPU memory, which is 40 GiB each. We report the simulation time of the largest experiments, which are Borealis's $M=216$ (high) and $M=288$ cases, Jiuzhang2.0's $M=144$ P65-5 case, and Jiuzhang3.0's $M=144$ (high) case. In all four cases, $\chi=10000$, $d=4$ for constructing the MPS, and $d=10$ for random displacement and sampling are used. Specifically, the longest time for the MPS calculation is 9.5 minutes for Jiuzhang2.0; other large experiments have similar times. The time for generating 10 million samples is 62 minutes.

\begin{figure*}[t]
\includegraphics[width=500px]{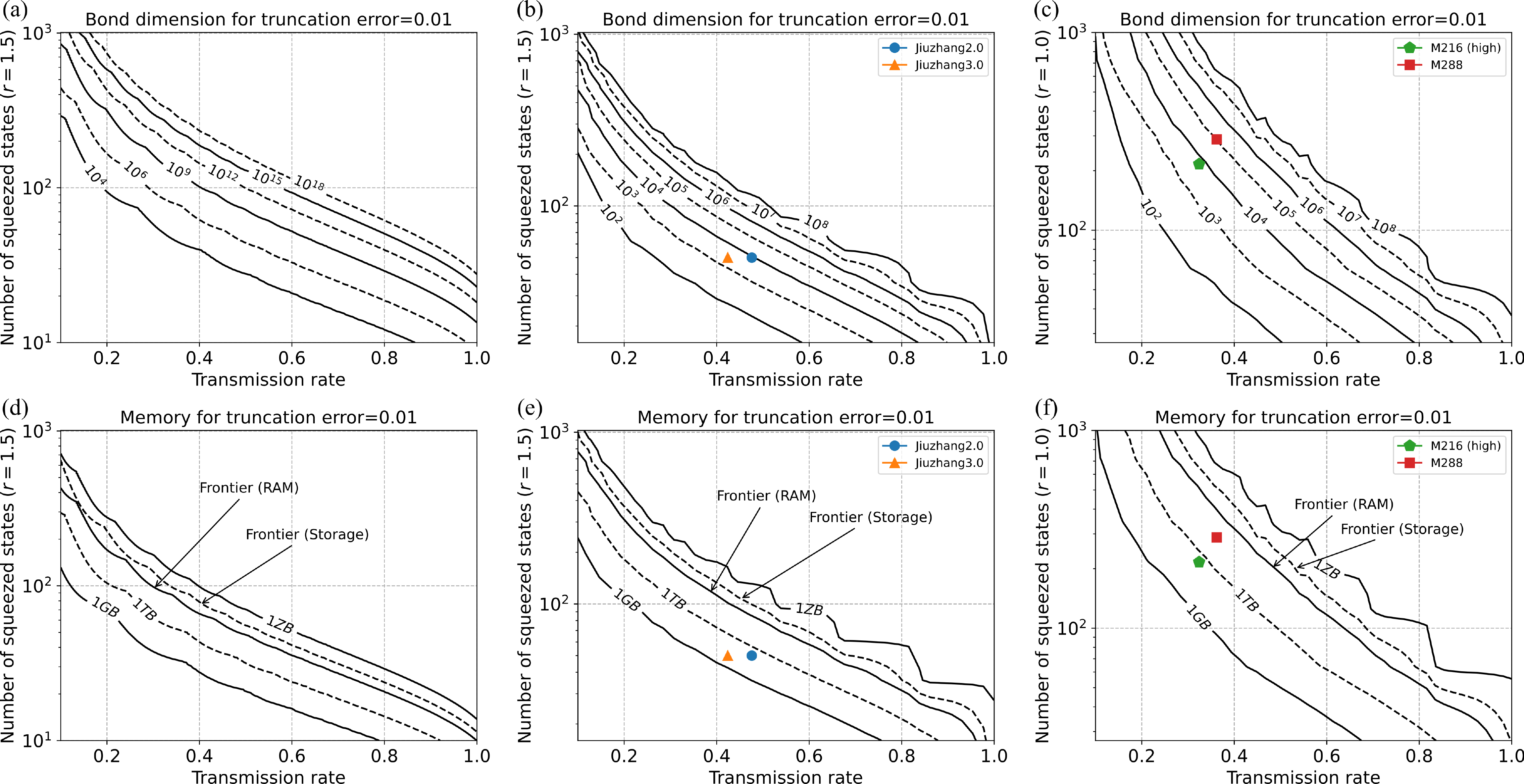}
\caption{Computational cost estimate to prepare an MPS for larger systems by (a)-(c) bond dimension of an MPS and (d)-(f) memory requirement to achieve truncation error $0.01$ with input squeezing parameter $r=1.5$. Here, (a) and (d) assume the worst-case circuit (see the main text) with squeezing parameter $r=1.5$, (b) and (e) assume a similar structure to the USTC's experiments with $r=1.5$, and (c) and (f) assume a similar structure to Xanadu's experiment with $r=1.0$.
Note that for consistency, we assume that all the input states are squeezed vacuum states with squeezing parameter $r$, i.e., the number of modes is equal to the number of squeezed states, which is not the same as the USTC's experiment.
The RAM and storage memory of Frontier, the current largest supercomputer, are $9.2$PB and $700$PB, respectively.
Here, we focus on the memory cost while we present the time cost in Appendix~\ref{app:estimate}.
We emphasize that the estimates are based on our algorithm, and there may exist more efficient classical algorithms.
We averaged over ten different random circuits to generate (b)(c)(e)(f).
}
\label{fig:estimate}
\end{figure*}

\section{Cost estimation for larger systems} \label{sec:estimate}
Finally, we estimate the computational cost of our algorithm to simulate larger systems to provide a guide for future quantum advantage experiments.
Before presenting the estimate, we emphasize that the estimate does not guarantee the boundary of quantum computational advantage because the estimate is based only on our current algorithm, and there may be possibilities to improve the algorithm further or find more efficient classical algorithms.

We focus on the cost for MPS preparation, which consumes larger costs than sampling for our simulation, and more particularly, on memory cost~(see Appendix~\ref{app:estimate} for more computational cost estimations, including time cost.)~and present the cost by the required bond dimension of an MPS and the storage memory to attain the MPS truncation error $0.01$, which is typical truncation error of our simulation as shown in Table~\ref{table:twopoint}.
Here, to make the analysis simple and consistent, we assume that~(i)~input squeezing parameters and loss rates are uniform across the modes,~(ii)~all the inputs are squeezed states without any empty inputs, i.e., vacuum states.
Note that the assumption~(i)~is not satisfied in Xanadu's experiments because the loss rate is not uniform for different paths of photons, and~(ii)~is not satisfied in the USTC experiment where some of the inputs are vacuum.
A more accurate estimate may depend on the specific configuration of experiments.
We consider three different circuit ensembles:~(a)~the worst-case circuits, which provides the maximum entanglement entropy as in Fig.~\ref{fig:bipartition},~(b)~the circuits similar to the USTC's experiments, and~(c)~the circuits similar to Xanadu's experiments.

Figure~\ref{fig:estimate} shows the estimate of the costs for different experimental parameters.
One can clearly see that different architectures require significantly different computational costs, highlighting the importance of detailed circuit configuration and circuit connectivity.
In particular, assuming that the storage capacity of Frontier, the current largest supercomputer, sets the boundary of classical computers, and for the worst-case circuit, if a transmission rate is $0.4$, we may need at least 80 squeezed states as an input. 
Or, if we use $50$ squeezed states, the transmission rate has to be larger than $0.538$.
For the USTC circuit type with the same loss rate, at least a hundred squeezed states are required for the memory cost to be comparable with Frontier's storage capacity.
For the Xanadu circuit type, at least five hundred squeezed states will be necessary.

Finally, it is worthwhile to emphasize again that the current estimates do not guarantee the boundary of quantum computational advantage because the estimates assume the particular algorithm, and there might exist possibilities of improving our algorithm significantly or developing other more efficient classical algorithms.
Moreover, the tolerable MPS truncation error may depend on other types of noise rates of each experiment, such as partial distinguishability, and how the tolerable truncation error scales as the system size needs to be further analyzed, incorporating the effect of the subsequent random displacement channel.
In addition, for some parameter regions, such as for low loss rates, other algorithms~\cite{quesada2022quadratic, bulmer2022boundary} might be a better option because our algorithm is particularly efficient for high loss cases.
Therefore, while our estimate will give a guide for future experiments, one has to analyze the cost in a more rigorous way with the associated parameters for different experimental setups.

\section{Discussion} \label{sec:discussion}
In this work, we have proposed a classical algorithm that can simulate the state-of-the-art experimental Gaussian boson samplers using moderate computational resources based on reasonable benchmarks.
One of the most important implications is that a large amount of photon loss is indeed significantly detrimental to the quantum computational advantage.
Therefore, for future experiments to rule out our classical algorithm to be implemented with reasonable resources, an obvious way is to significantly improve the transmission rate rather than increase the input squeezing parameters.
In fact, many experiments have focused on increasing the input squeezing parameters to increase the output photon numbers.
However, our algorithm manifestly shows that the actual squeezing parameters are more important to increase the complexity, which will guide the direction of future experiments for achieving larger complexity.
Also, when the transmission rate is inevitably limited, increasing the number of squeezed states  makes our algorithm's complexity exponentially increase, which can be another path to overcome our algorithm's computational power.
In addition, as observed in our numerical simulation, the circuit connectivity is another crucial factor to make the tensor network approach intractable by increasing the entanglement.
Furthermore, the tolerable error for an MPS can significantly change the required bond dimension and thus the computational cost.
Therefore, reducing the other types of noise is also crucial to increase the complexity.

Another interesting implication of our algorithm is that as the random circuit sampling case~\cite{morvan2023phase}, the complexity of experimental Gaussian boson sampling can now be characterized by an MPS's bond dimension.
Therefore, it allows us to compare the computational complexity of a random circuit sampling experiment and a Gaussian boson sampling experiment.

Moreover, aside from the context of Gaussian boson sampling for quantum advantage, our algorithm may be beneficial for other applications.
A prominent application is a simulation for non-Gaussian state preparation using lossy Gaussian boson sampling with post-selection, such as Gottesman-Kitaev-Preskill state and cat state~\cite{quesada2019simulating, su2019conversion}, which are important resources for various quantum information processing.

We now present open questions that were not addressed in this work.
First, an obvious question is whether there is a better classical algorithm than the presented one for current-size Gaussian boson sampling or in an asymptotic regime.
Regarding the latter, whereas our algorithm provides a significant advantage for fixed-size experiments, our algorithm does not render a particular advantage over the existing algorithms asymptotically~\cite{oszmaniec2018classical, garcia2019simulating, qi2020regimes}.
Finding a clear boundary between an easy and hard regime in terms of loss scaling is still an open question.
Also, our entropic analysis of the algorithm's complexity is based only on the quantum part $V_p$.
However, the additional random displacement channel may decrease the distance between the algorithm's output distribution and the ground-truth distribution due to the contractivity of trace distance over quantum channels.
Thus, another important open question is incorporating the random displacement channel's effect on the simulation complexity and accuracy.
Moreover, our algorithm exploits many properties of Gaussian states.
Generalization of our algorithm to more general cases, such as Fock-state boson sampling~\cite{aaronson2011computational}, is another interesting open question.

\section{Methods}\label{sec:methods}
\subsection{Covariance matrix formalism}\label{methods:cov}
Let us introduce the covariance matrix formalism~\cite{ferraro2005gaussian, weedbrook2012gaussian, serafini2017quantum}.
An $M$-mode Gaussian state $\hat{\rho}$'s covariance matrix of Wigner function is given by $2M\times 2M$ matrix $V$, whose elements are written as $V_{ij}=\langle \hat{q}_i\hat{q}_j+\hat{q}_j\hat{q}_i\rangle/2-\langle \hat{q}_i\rangle\langle\hat{q}_j\rangle$, where $\langle \hat{O} \rangle=\Tr[\hat{\rho}\hat{O}]$.
Here, $\hat{q}=(\hat{x}_1,\dots,\hat{x}_M,\hat{p}_1,\dots,\hat{p}_M)$ is the quadrature operator vector with position and momentum quadrature operators $\hat{x}_i$'s and $\hat{p}_i$'s. 
When a given quantum state is Gaussian and its mean vector of quadratures is zero, the covariance matrix $V$ can uniquely identify the quantum state.
The condition that a physical covariance matrix has to satisfy, namely, that its density matrix is positive semidefinite, is the canonical commutation relation
\begin{align}
    V\succeq i\Omega,
    ~~\text{where}~
    \Omega\equiv\omega \otimes \mathbb{1}_M,
    ~~
    \omega\equiv
    \begin{pmatrix} 0 & 1 \\ -1 & 0
    \end{pmatrix}.
\end{align}
For a given covariance matrix, we can obtain the average photon number by $\Tr[V-\mathbb{1}_{2M}]/4$ if the first-moment vector of the state is zero.

\subsection{Efficiency of Matrix product state}\label{methods:mps}
Here, we present more details about the efficiency of MPS representation, i.e., when an efficient MPS representation exists; that is, the required bond dimension is $\chi=\poly(K,1/\epsilon)$, where $\epsilon$ is the trace distance between the true state and the approximate MPS~and $K$ is the system size.
It is well-known from Refs.~\cite{verstraete2006matrix, schuch2008entropy} that in order to prove that a given quantum state's MPS approximation is efficient, it suffices to investigate the R{\'e}nyi entropy of its reduced density matrix obtained by bipartition.
More specifically, the state allows an efficient MPS representation when the R{\'e}nyi entropy entanglement with $\alpha<1$ increases at most logarithmically with the system size. 
On the other hand, when the R{\'e}nyi entropy entanglement with $\alpha>1$ increases at least $\Omega(K^\gamma)$ with the system size $K$ and a constant $\gamma$, the state does not allow an efficient MPS representation.

For a large loss rate, that is, where the loss rate converges to zero as the system size grows $\eta\to 0$, ($\eta=O(K^{-\beta})$ with $\beta>0$), we have
\begin{align}
    s=-\frac{1}{2}\log\left(\eta e^{-2r}+1-\eta\right)
    \approx \eta e^{-r}\sinh{r}+O(\eta^2).
\end{align}
In this case, the R{\'e}nyi entropy for a two-mode squeezed state is given by
\begin{align}
    S_\alpha(s)
    =\frac{\log\Tr[\hat{\rho}_A^\alpha]}{1-\alpha}
    =\frac{\log (\cosh^{2\alpha}s-\sinh^{2\alpha}s)}{\alpha-1}.
\end{align}
Thus, for $K$ two-mode squeezed states, the R{\'e}nyi entropy with $\alpha<1$ and $\eta\to0$ becomes
\begin{align}
    KS_\alpha(s)
    =O(K\eta^{2\alpha}).
\end{align}
On the other hand, when $\alpha=1.01$, the total R{\'e}nyi entropy is written as
\begin{align}
    KS_\alpha(s)
    =\Omega(K\eta^{2}).
\end{align}

\subsection{Cross-entropy benchmarking (XEB)}\label{methods:XEB}
Here, we formally define the XEB, which is frequently used as a benchmark for sampling.
For Gaussian boson sampling specifically, XEB is defined for each photon sector $N$ as
\begin{align}
    \text{XE}_N=\frac{1}{N_s}\sum_{i=1}^{N_s} \log \frac{p_\text{id}(S_i)}{\mathcal{N}},~~~ \mathcal{N}=\frac{\text{Pr}(N)}{\binom{N+M-1}{N}},
\end{align}
where $N_s$ is the number of samples, $S_i$'s are the $i$th sample's photon number pattern whose total photon number is $N$, $p_\text{id}(S_i)$ is the ground-truth probability of obtaining $S_i$, and $\text{Pr}(N)$ is the probability of obtaining total $N$ photons from the ground-truth distribution.
For Jiuzhang's cases, where threshold detectors are used, the normalization factor is replaced by $\mathcal{N}=\Pr(N)/\binom{M}{N}$, where $\Pr(N)$ is the probability of obtaining total $N$ photon clicks from the ground-truth distribution.

\section{Data Availability}
Samples generated from our method can be found at \doi{10.17605/OSF.IO/49TRH}.

\section{Code Availability}
Code for our numerical simulation and data analysis can be found at \doi{10.5281/zenodo.7992736}.

\appendix

\section{Existing classical algorithms and our classical algorithm}\label{app:comparison}
In this appendix we compare our classical algorithm with existing algorithms simulating Gaussian boson sampling.
First, we compare it with the {\it exact} simulation algorithms from Refs.~\cite{quesada2020exact, quesada2022quadratic, bulmer2022boundary}.
These exact simulation algorithms are typically used for estimating the running time for classical computers to simulate the experiments~\cite{zhong2020quantum, zhong2021phase, madsen2022quantum}. 
One significant limitation of such algorithms is that their complexity increases substantially even if we simply introduce a large displacement before the photon number measurement.
This is because the large displacement will introduce more output photons on average, which increases the matrix size of which we compute the (loop) hafnian for sampling.
It implies that the classical algorithms do not discriminate between quantum resources, which are photons from squeezed states, and classical resources, which are photons from displacement, as long as there exist quantum resources.
Therefore, even when there are a large number of thermal photons due to photon loss, which is the case in the current experiments (see Table~\ref{table:parameters}), the complexity analysis based on these types of algorithms inevitably overestimates the complexity of the corresponding experiments.
Furthermore, the appropriate target of hardness is typically an approximate simulation instead of an exact simulation~\cite{aaronson2011computational, hamilton2017gaussian, deshpande2021quantum}.
Therefore, the running time from these exact algorithms overestimates the complexity of the existing experiments, taking into account additional experimental noise.



Let us now introduce {\it approximate} classical algorithms designed to simulate {\it lossy} boson sampling.
Various types of algorithms exist for this purpose.
First, we consider efficient classical algorithms using classical states.
The key idea of these algorithms is to find the closest classical state, such as a thermal state~\cite{garcia2019simulating, qi2020regimes, martinez2023classical} or a separable state~\cite{oszmaniec2018classical}, which are easy to efficiently sample from.
One significant drawback of these algorithms is that they cannot improve the accuracy by increasing the running time because the closest classical state is fixed for each algorithm for given parameters of a lossy boson sampler, and thus we cannot enhance the quality of the approximate state by increasing the running time.
Therefore, although these approximations provide a boundary of efficient classical sampling and an important benchmark for demonstrating quantum advantage, they are not appropriate to simulate lossy boson samplers or characterize their complexity unless the loss rate is very large.

Another relevant type of classical algorithm that takes advantage of photon loss is matrix product operators~\cite{huang2019simulating, oh2021classical, liu2023simulating}.
The main idea is that since external noise, such as photon loss, reduces the amount of the entanglement of the output state, such a tensor network method may be able to reduce the complexity of simulating the corresponding lossy boson samplers.
However, although  the required bond dimension does not necessarily increase rapidly for the high loss regime, simulating the recent experiments is still too expensive as the output photon number increases.
Another drawback is that since the matrix product operators simulate density matrices, the cost is at least quadratically larger than simulating a pure state, which is the case in the present work.

Our algorithm addresses such drawbacks of the existing algorithms.
Although our algorithm is generally an approximate sampler because of the truncation of singular values, the complexity of our algorithm does not increase by introducing a displacement on the output state.
Therefore, one can significantly reduce the complexity when a given boson sampler contains a large number of thermal photons.
In addition, our algorithm can efficiently improve its accuracy from a thermal-state approximation to the ideal distribution by increasing the running time depending on the target approximation error, which is due to the fact that the MPS method enables us to take into account the contribution from $V_p$ in contrast to the thermal state approximation that approximates $V_p$ by a trivial vacuum state's covariance matrix $\mathbb{1}_{2M}$.
As emphasized repeatedly in the main text, an interesting feature of our algorithm is that in one extreme case where we set the MPS part to be a vacuum state, it becomes a thermal state approximation, which recovers the concept of the previous results in Refs.~\cite{garcia2019simulating, qi2020regimes}
Also, our new algorithm has a significant advantage over the previous matrix product operator algorithms~\cite{oh2021classical, liu2023simulating} in that we remove the thermal photons $W$, which enables a drastic reduction in resource utilization for the MPS algorithm.

At a high level, our algorithm resembles another type of classical algorithm designed for approximate noisy samplers exploiting the fact that the output probability of noisy quantum devices typically converges to an easy distribution~\cite{kalai2014gaussian, renema2018efficient, renema2018classical, shchesnovich2019noise, moylett2019classically, villalonga2021efficient, aharonov2022polynomial, oh2023classical}.
More formally, the noisy output probabilities can be expanded as polynomials with different degrees in terms of the noise parameter, with the lowest degree of polynomials being a distribution that is easy to sample from (e.g., uniform distribution in Refs.~\cite{aharonov2022polynomial, oh2023classical}). 
Thus, by appropriately choosing the degree, one can control the accuracy of the approximate samplers by increasing the running time.
Our approximate sampler resembles this type of algorithm in that one extreme case gives a distribution that is easy to sample from, and as we increase the running time, we can improve the algorithm's accuracy.
To the best of our knowledge, however, this type of classical algorithm requires a large amount of noise to be efficient and so has not been implemented for finite-size experiments yet.

Still another type of classical algorithm  is specialized to spoof benchmarking but not necessarily to simulate the ground-truth sampler~\cite{oh2023spoofing}.
We emphasize that the proposed algorithm in this work is not designed to spoof benchmarking but truly to approximately simulate the ground-truth distribution.

\section{Single-mode decomposition}\label{app:single}
In this appendix we consider the decomposition of the simplest example, a single-mode Gaussian state with photon loss.
Suppose we have prepared a squeezed vacuum state with squeezing parameter $r\geq 0$, which we assumed nonnegative without loss of generality, whose covariance matrix is given by
\begin{align}
    V_0=
    \begin{pmatrix}
        e^{2r} & 0 \\
        0 & e^{-2r}
    \end{pmatrix}.
\end{align}
After the photon loss channel, whose transmission rate is given by $\eta$ and loss rate is thus  $1-\eta$, the covariance matrix transforms to
\begin{align}
    V=
    \eta V_0+(1-\eta)\mathbb{1}_2
    =
    \begin{pmatrix}
        \eta e^{2r}+1-\eta & 0 \\
        0 & \eta e^{-2r}+1-\eta
    \end{pmatrix}.
\end{align}
By setting $\eta e^{-2r}+(1-\eta)=e^{-2s}$, we can decompose the covariance matrix $V$ as
\begin{align}
    V=
    \begin{pmatrix}
        e^{2s} & 0 \\
        0 & e^{-2s}
    \end{pmatrix}
    +
    \begin{pmatrix}
        \eta e^{2r}+1-\eta-e^{2s} & 0 \\
        0 & 0
    \end{pmatrix}
    \equiv V_p+W,
\end{align}
where $W$'s first matrix element is nonnegative when $r\geq 0$.
Here we call $s$ an actual squeezing parameter because this is the squeezing parameter that is actually involved in complexity, which will be clear below.
We note that such a decomposition is not unique.
For our purpose, however, we maximize the matrix $W$ and minimize the energy of the pure Gaussian state $V_p$ to minimize the quantum resource (see below for another decomposition from the Williamson decomposition.).
We emphasize that a similar decomposition has been used to accelerate the classical algorithm for simulating Gaussian boson sampling quadratically in Ref.~\cite{quesada2022quadratic}.
However, the purpose of the decomposition was to avoid simulating the density matrix rather than a pure state, not to take advantage of the fact that the thermal part, namely, the random displacement part, should not significantly increase the complexity.
Furthermore, the optimization we implement  significantly  minimizes the simulation cost.
As an example, the Williamson decomposition from Ref.~\cite{quesada2022quadratic} gives the following decomposition:
\begin{align}
    V&=
    \begin{pmatrix}
        e^{t} & 0 \\
        0 & e^{-t}
    \end{pmatrix}
    \begin{pmatrix}
        2n_\text{nth}+1 & 0 \\
        0 & 2n_\text{nth}+1
    \end{pmatrix}
    \begin{pmatrix}
        e^{t} & 0 \\
        0 & e^{-t}
    \end{pmatrix} \\ 
    &=
    \begin{pmatrix}
        e^{t} & 0 \\
        0 & e^{-t}
    \end{pmatrix}
    \left[
    \begin{pmatrix}
        1 & 0 \\
        0 & 1
    \end{pmatrix}
    +
    2\begin{pmatrix}
        n_\text{th} & 0 \\
        0 & n_\text{th}
    \end{pmatrix}
    \right]
    \begin{pmatrix}
        e^{t} & 0 \\
        0 & e^{-t}
    \end{pmatrix} \\ 
    &=
    \begin{pmatrix}
        e^{2t} & 0 \\
        0 & e^{-2t}
    \end{pmatrix}
    +
    2\begin{pmatrix}
        n_\text{th}e^{2t} & 0 \\
        0 & n_\text{th}e^{-2t}
    \end{pmatrix}.
\end{align}
Here, the squeezing parameter of the pure state is
\begin{align}
    t=\frac{1}{4}\log\left(\frac{\eta e^{2r}+1-\eta}{\eta e^{-2r}+1-\eta}\right),
\end{align}
and
\begin{align}
    n_\text{th}=\frac{1}{2}\left(\sqrt{(\eta e^{2r}+1-\eta)(\eta e^{-2r}+1-\eta)}-1\right).
\end{align}
One can easily show that $t>s$, namely, our minimization, gives a smaller photon number for the quantum part.

\section{MPS construction}\label{app:MPS}
In this appendix we provide the method for MPS construction of Gaussian states based on Ref.~\cite{vidal2003efficient}.
While the method in that reference is typically inefficient, we employ the property of Gaussian states to make it efficient so that we can efficiently find the reduced density matrix of a bipartition $A:B$ and its spectral decomposition because the reduced density matrix is still a Gaussian state and Gaussian states can always be written as a product of thermal states $\hat{\rho}_T$ followed by a Gaussian unitary operation~\cite{serafini2017quantum}:
\begin{align}\label{eq:decompose_G}
    \hat{\rho}_B
    =\Tr_A[|\psi\rangle\langle\psi|]
    =\hat{U}_B \hat{\rho}_T\hat{U}_B^\dagger
    =\sum_{\bm{n}=\bm{0}}^\infty p_T(\bm{n})\hat{U}_B |\bm{n}\rangle\langle \bm{n}|\hat{U}_B^\dagger,
\end{align}
where $p_T(\bm{n})=\prod_{i\in B}\bar{n}_i^{n_i}/(\bar{n}_i+1)^{n_i+1}$ and $\bar{n}_i$ is the mean photon number of the $i$th mode's thermal state.
One can easily find $\hat{U}_B$ and $\{\bar{n}_i\}_{i\in B}$ by Williamson decomposition of the covariance matrix of the state in the $B$ part~\cite{serafini2017quantum}.
Hence, the eigenstates of the reduced density matrix are always a number state followed by a Gaussian unitary operation.

Here we recall the method of constructing an MPS proposed in Ref.~\cite{vidal2003efficient} with adapting Gaussian states' properties.
First, we apply the singular value decomposition along the first mode and the rest of the modes with a prechosen bond dimension $\chi$:
\begin{align}
    |\psi\rangle
    &\approx \sum_{\alpha_1=0}^{\chi-1}\lambda_{\alpha_1}^{[1]}|\Phi_{\alpha_1}^{[1]}\rangle|\Phi_{\alpha_1}^{[2\cdots M]}\rangle \\ 
    &=\sum_{n_1=0}^{d-1}\sum_{\alpha_1=0}^{\chi-1}\Gamma_{\alpha_1}^{[1]n_1}\lambda_{\alpha_1}^{[1]}|n_1\rangle|\Phi_{\alpha_1}^{[2\cdots M]}\rangle \\
    &=\sum_{n_1=0}^{d-1}\sum_{\alpha_1=0}^{\chi-1}A_{\alpha_1}^{[1]n_1}|n_1\rangle|\Phi_{\alpha_1}^{[2\cdots M]}\rangle,
\end{align}
where
\begin{align}
    A_{\alpha_1}^{[1]n_1}
    &=\langle n_1^{[1]}|\langle \Phi_{\alpha_1}^{[2\cdots M]}|\psi\rangle \\
    &=\langle n_1^{[1]}|\langle \bm{n}_{\alpha_1}^{[2\cdots M]}|\hat{U}^{[2\cdots M]\dagger}\hat{U}|\bm{n}=\bm{0}\rangle.
\end{align}
Here the approximation is due to truncation from the predetermined bond dimension, and the state $|\psi\rangle$ can always be written as $|\psi\rangle=\hat{U}|\bm{n}=\bm{0}\rangle$ by Williamson decomposition of a pure Gaussian state; and, as emphasized before, the singular values $\lambda_{\alpha_1}^{[1]}$ can be easily found by performing the Williamson decomposition for the marginal covariance matrix over the bipartition between the first mode and the rest of the modes as in Eq~\eqref{eq:decompose_G}.
Also, the eigenstates $\{|\Phi_{\alpha_1}^{[1]}\rangle\}$, $\{|\Phi_{\alpha_1}^{[2\cdots M]}\rangle\}$ are always photon number states followed by Gaussian unitary operations.
Thus, we can characterize the eigenstate $|\Phi_{\alpha_1}^{[2\dots M]}\rangle=\hat{U}^{[2\cdots M]}|\bm{n}_{\alpha_1}^{[2\cdots M]}\rangle$ as $\{\hat{U}^{[2\cdots M]},\bm{n}_{\alpha_1}^{[2\cdots M]}\}$.
We then rewrite it in number basis $|n_2^{[2]}\rangle$ as
\begin{align}
    |\Phi_{\alpha_1}^{[2\cdots M]}\rangle
    =\sum_{n_2=0}^{d-1}|n_2^{[2]}\rangle|\tau_{\alpha_1n_2}^{[3\cdots M]}\rangle,
\end{align}
where
\begin{align}
    |\tau_{\alpha_1n_2}^{[3\cdots M]}\rangle
    =\langle n_2^{[2]}|\Phi_{\alpha_1}^{[2\cdots M]}\rangle,
\end{align}
where we expand by the eigenstates of the reduced density matrix $\{|\Phi_{\alpha_2}^{[3\cdots M]}\rangle\}_{\alpha_2=0}^{\chi-1}$
\begin{align}
    |\tau_{\alpha_1n_2}^{[3\cdots M]}\rangle
    &\approx\sum_{\alpha_2=0}^{\chi-1}A_{\alpha_1\alpha_2}^{[2]n_2}|\Phi_{\alpha_2}^{[3\cdots M]}\rangle
    =\sum_{\alpha_2=0}^{\chi-1}\Gamma_{\alpha_1\alpha_2}^{[2]n_2}\lambda_{\alpha_2}^{[2]}|\Phi_{\alpha_2}^{[3\cdots M]}\rangle \\ 
    A_{\alpha_1\alpha_2}^{[2]n_2}
    &=\langle n_2^{[2]}|\langle \Phi_{\alpha_2}^{[3\cdots M]}|\Phi_{\alpha_1}^{[2\cdots M]}\rangle,
\end{align}
where $A_{\alpha_1\alpha_2}^{[2]n_2}=\Gamma_{\alpha_1\alpha_2}^{[2]n_2}\lambda_{\alpha_2}^{[2]}$, and $|\Phi_{\alpha_2}^{[3\cdots M]}\rangle$ is the eigenstate of the reduced density matrix of the $[3\cdots M]$ part and $\lambda_{\alpha_2}^{[2]}$ are the singular values, which can be easily identified.
Practically, we  need only to compute matrices $A$ by
\begin{align}
    A_{\alpha_1\alpha_2}^{[2]n_2}
    &=\langle n_2^{[2]}|\langle \Phi_{\alpha_2}^{[3\cdots M]}|\Phi_{\alpha_1}^{[2\cdots M]}\rangle \\
    &=\langle n_2^{[2]}|\langle \bm{n}_{\alpha_2}^{[3\cdots M]}|\hat{U}^{[3\cdots M]\dagger}\hat{U}^{[2\cdots M]}|\bm{n}_{\alpha_1}^{[2\cdots M]}\rangle \\
    &=\langle n_2^{[2]}|\langle \bm{n}_{\alpha_2}^{[3\cdots M]}|\hat{V}^{[2\cdots M]}|\bm{n}_{\alpha_1}^{[2\cdots M]}\rangle,
\end{align}
where $\hat{V}^{[2\cdots M]}\equiv \hat{U}^{[3\cdots M]\dagger}\hat{U}^{[2\cdots M]}$.
By iterating this procedure, we obtain all the matrix elements, which is summarized as
\begin{align}
    &A_{\alpha_1}^{[1]n_1}
    =\langle n_1^{[1]}|\langle \bm{n}_{\alpha_1}^{[2\cdots M]}|\hat{U}^{[2\cdots M]\dagger}\hat{U}^{[1\cdots M]}|\bm{n}=\bm{0}\rangle, \\ 
    &A_{\alpha_{k-1}\alpha_k}^{[k]n_k}
    =\langle n_k^{[k]}|\langle \bm{n}_{\alpha_k}^{[(k+1)\cdots M]}|\hat{U}^{[(k+1)\cdots M]\dagger}\hat{U}^{[k\cdots M]}|\bm{n}_{\alpha_{k-1}}^{[k\cdots M]}\rangle \nonumber\\
    &~~~~~~~~~~~~~~~~~~~~~~~~~~~~~~~~~~~~~~~~~~~~\text{for}~~ 1<k<M, \\ 
    &A_{\alpha_{M-1}}^{[M]n_M}
    =\langle n_M^{[M]}|\hat{U}^{[M]}|n_{\alpha_{M-1}}^{[M]}\rangle,
\end{align}
and 
\begin{align}
    &\Gamma^{[1]n_1}_{\alpha_1}=A^{[1]n_1}_{\alpha_1}/\lambda^{[1]}_{\alpha_1},\\
    &\Gamma^{[k]n_k}_{\alpha_{k-1}\alpha_{k}}=A^{[k]n_k}_{\alpha_{k-1}\alpha_{k}}/\lambda^{[k]}_{\alpha_k}~~ \text{for}~~ 1<k<M, \\
    &\Gamma^{[M]n_M}_{\alpha_{M-1}}=A^{[M]n_{M}}_{\alpha_{M-1}}/\lambda^{[M-1]}_{\alpha_{M-1}}.    
    \end{align}

Therefore, the remaining calculation to obtain all the matrix elements is $\langle \bm{n}_1|\hat{V}|\bm{n}_2\rangle$, for number states $|\bm{n}_1\rangle$ and $|\bm{n}_2\rangle$ and a Gaussian unitary operator $\hat{V}$.
This quantity has already been studied in Refs.~\cite{quesada2019franck, oh2022quantum} by noting that any Gaussian unitary operation can be decomposed as $\hat{V}=\hat{U}_2\hat{S}(\bm{r})\hat{U}_1$, where passive unitary operations $\hat{U}_1$ and $\hat{U}_2$ and single-mode squeezers $\hat{S}(\bm{r})=\otimes_{i}\hat{S}(r_i)$:
\begin{align}
    \langle \bm{n}_1|\hat{U}_2\hat{S}(\bm{r})\hat{U}_1|\bm{n}_2\rangle
    =\frac{\haf(\Sigma_{\bm{n}_1,\bm{n}_2})}{\sqrt{\bm{n}_1!\bm{n}_2!\prod_i \cosh{r_i}}},
\end{align}
where $U_2$ and $U_1$ correspond to the unitary matrix that characterize the unitary operators $\hat{U}_2$ and $\hat{U}_1$, and $\Sigma$ is a matrix obtained by
\begin{align}
    \Sigma=
    \begin{pmatrix}
        U_2 & 0 \\ 
        0 & U_1^\text{T}
    \end{pmatrix}
    \begin{pmatrix}
        \tanh{\bm{r}} & \sech{\bm{r}} \\ 
        \sech{\bm{r}} & -\tanh{\bm{r}}
    \end{pmatrix}
    \begin{pmatrix}
        U_2^\text{T} & 0 \\ 
        0 & U_1
    \end{pmatrix},
\end{align}
where $\Sigma_{\bm{n}_1,\bm{n}_2}$ is obtained by repeating $\Sigma$'s block matrices by $\bm{n}_1$ and $\bm{n}_2$ times.
Hence, the complexity of obtaining all the matrix elements of $A$, or equivalently $\Gamma$ and $\lambda$, is $O(Md\chi^2\times (\text{hafnian of $\Sigma_{\bm{n}_1,\bm{n}_2}$}))$, and the complexity of computing the hafnian of $\Sigma_{\bm{n}_1,\bm{n}_2}$ is $\tilde{O}(2^{(\bm{n}_1+\bm{n}_2)/2})$~\cite{bjorklund2012counting, bjorklund2019faster}.
Therefore, two crucial factors determine the complexity: the bond dimension $\chi$ and the maximum of $|\bm{n}_1+\bm{n}_2|\equiv \sum_{i}((\bm{n}_1)_i+(\bm{n}_2)_i)$.
While we study the scaling of the bond dimension $\chi$ in Sec.~\ref{sec:entropy} more comprehensively, both the bond dimension and the maximum $|\bm{n}_1+\bm{n}_2|$ are affected by the amount of entanglement.
The former is evident, and the latter is because for a pure multimode Gaussian state, the reduced state on part $B$ over a bipartition $A:B$ becomes more thermalized when the parties $A$ and $B$ are highly entangled.
For example, if they are a product state, the reduced state is still a pure state.
Also, we emphasize that the matrix size of $\Sigma_{\bm{n}_1,\bm{n}_2}$ is much smaller than the matrix size for computing the output probability of the actual output state, which includes the random displacement.
Hence, our MPS construction is, in general, much more efficient than directly sampling from the output state using the best-known classical algorithm~\cite{bulmer2022boundary, quesada2022quadratic} when the loss rate is large.

\begin{widetext}
\section{Singular values of output state of Gaussian boson sampling}\label{app:error}
In this appendix we analyze the distribution of singular values of MPS more quantitatively.
By analysis, we show that we  need only the bond dimension $\chi=\poly\log(1/\epsilon)$ for a fixed number of squeezed states and squeezing parameters and that we may not efficiently reduce the simulation error for the transmission rate scaling as $\eta=O(1/\sqrt{N})$.
To this end, we consider $K$ two-mode squeezed states out of $M$ mode for fixed $K$ and $M$ and provide a way to truncate singular values as in Sec.~\ref{sec:entropy}.
We directly study $K$ two-mode squeezed states' singular values for a bipartition instead of entanglement entropy.
Then the reduced density matrix of one part of two-mode squeezed states can be written as a product of thermal states
\begin{align}
    \hat{\rho}_T(\bar{n})^{\otimes K}
    =\bigotimes_{i=1}^K\left(\sum_{n_i=0}^\infty p_{\bar{n}}(n_i)|n_i\rangle\langle n_i|\right)
    =\sum_{\bm{n}=0}^\infty p_{\bar{n}}(\bm{n})|\bm{n}\rangle\langle \bm{n}|
    =\sum_{k=0}^\infty \sum_{\bm{n}:\sum_i n_i=k}(1-R)^K R^k|\bm{n}\rangle\langle \bm{n}|,
\end{align}
where $\bar{n}=\sinh^2 s$ is the mean photon number of thermal states $\hat{\rho}_T$ obtained by partial trace, $p_{\bar{n}}(n_i)\equiv\bar{n}_i^{n_i}/(\bar{n}_i+1)^{n_i+1}$, and $p_{\bar{n}}(\bm{n})\equiv \prod_{i=1}^Kp_{\bar{n}}(n_i)$.
Here we have decomposed the sum into the total photon number sector $k$ and outcomes that have the same photon number, namely, $\bm{n}$ such that $\sum_i n_i=k$.
For the last expression, we define $R\equiv \bar{n}/(\bar{n}+1)$
Then for each sector $k$, we have eigenvalues $(1-R)^K R^k$ for $\binom{K+k-1}{k}$ terms.
Thus, the distribution of the sum of eigenvalue for each sector $k$ follows a negative binomial distribution.
Now we analyze the error when we truncate this up to $l$ photon sector.
For this case the number of singular values we keep is given by
\begin{align}
    \chi_l\equiv \sum_{k=0}^l\binom{K+k-1}{k},
\end{align}
which becomes the bond dimension we use for MPS.
The probability we lose by the truncation is
\begin{align}\label{eq:truncation}
    \epsilon_l\equiv \sum_{k=l+1}^\infty \binom{K+k-1}{k}(1-R)^K R^k
    =1-I_{1-R}(K,l+1),
\end{align}
where $I_R$ is regularized incomplete beta function.
Here, since $(1-R)^K R^k$ is a monotonically decreasing function in $k$, the best way of choosing the singular values to minimize the error is to choose from the lowest photon number sectors.
We also note that the truncation number $l$ determines the largest $|\bm{n}_1+\bm{n}_2|$ that appeared in the direction construction of MPS of the output Gaussian state (see Sec.~\ref{sec:mps}).
Hence, if $l$ needs to increase at most logarithmically, the part for computing hafnians for the direct construction of the MPS is efficient.

Note that the tail probability of binomial distribution with probability $p$ of obtaining at least $k+1$ success out of $n$ trials, or, equivalently, multiple Bernoulli trials, is $1-I_{1-p}(n-k,1+k)$.
Thus, the right-hand side is equivalent to obtaining at least $l+1$ successes from the binomial with $K+l$ trials with success probability $R$.
Using concentration inequality for $X\equiv \sum_{i=1}^{K+l}X_i$ with Bernoulli random variable $X_i$ with probability $R$~\cite{dubhashi2009concentration}, we have
\begin{align}
    \Pr[X>(1+\delta)\mathbb{E}(X)]\leq \exp\left(-\frac{\delta^2}{3}\mathbb{E}[X]\right),
\end{align}
and choosing $\delta=l/\mathbb{E}[X]-1=l/[(K+l)R]-1>0$ (thus, $l>KR/(1-R)$),
the probability we lose by truncation is upper bounded as
\begin{align}
    \epsilon_l
    =1-I_{1-R}(K,l+1)
    =\Pr[X>l]
    \leq \exp\left(-\frac{1}{3}\left(\frac{l}{(K+l)R}-1\right)^2(K+l)R\right)
    \leq \exp\left(-\frac{1}{3}c^2lR\right),
\end{align}

Now let us consider a fixed-size circuit, that is, one in which $K$, $\bar{n}$, and $R$ are fixed.
Then
\begin{align}
    \epsilon_l\leq \exp\left(-\frac{1}{3}\left(\frac{l}{(K+l)R}-1\right)^2(K+l)R\right)
    \leq \exp\left(-\frac{1}{3}c^2lR\right),
\end{align}
where $c$ is a nonzero constant lower bound of $\delta$ for $l>KR/(1-R)$.
Thus, in order to bound the error by $\epsilon$, it is sufficient to choose
\begin{align}
    l
    \geq \frac{-3\log \epsilon}{c^2 R}
    =O(\log\epsilon^{-1}),
\end{align}
which leads to the upper bound of the bond dimension
\begin{align}
    \chi_l=\sum_{k=0}^l \binom{K+k-1}{k}
    \leq (l+1)\binom{K+l-1}{l}
    \leq (l+1)\frac{(K+l-1)^{K-1}}{(K-1)!}
    =O(l^K)
    =O\left(\left(\log \epsilon^{-1}\right)^{K}\right).
\end{align}
Therefore, the bond dimension to achieve error $\epsilon$ is $\chi=\poly\log(\epsilon^{-1})$.
It also guarantees that the computation of hafnians is efficient for the direct construction of MPS since $l$ can be chosen to be $O(\log \epsilon^{-1})$.

Now let us turn our attention to the case where $\eta=\Theta(1/\sqrt{K})$.
When $\eta=\Theta(1/\sqrt{K})$, $s=\Theta(1/\sqrt{K})$, $\bar{n}=\Theta(1/K)$, and $R=\bar{n}/(\bar{n}+1)\approx 1/K$.
First, we note that if we select $l=0$, the number of singular values is $\chi_0=1$ and the error is also $\epsilon_0=1-(1-R)^K\approx 1-(1-1/K)^K\approx 1-1/e$.
Thus, it recovers the same scaling as Refs.~\cite{oszmaniec2018classical, garcia2019simulating,qi2020regimes}.

Again, using concentration inequality for $X\equiv \sum_{i=1}^{K+l}X_i$ with Bernoulli random variable $X_i$ with probability $R$ and choosing $\delta=l/\mathbb{E}[X]-1=l/[(K+l)R]-1>0$ (thus, $l>KR/(1-R)$), we have
\begin{align}
    \epsilon_l
    \leq \exp\left(-\frac{1}{3}\left(\frac{l}{(K+l)R}-1\right)^2(K+l)R\right).
\end{align}
The difference from the fixed-size circuit case is that we are interested in the regime where $R\approx \bar{n}\approx 1/K$.
Note that we chose $l$ to be larger than $KR/(1-R)=K\bar{n}=O(1)$ for $R=O(1/K)$.
Then the bound we obtained gives, for large $K$,
\begin{align}
    \exp\left(-\frac{1}{3}\left(\frac{l}{(K+l)R}-1\right)^2(K+l)R\right)
    \approx \exp\left(-\frac{1}{3}\left(\frac{Kl}{(K+l)}-1\right)^2\frac{K+l}{K}\right)
    \approx \exp\left(-\frac{1}{3}\left(l-1\right)^2\right),
\end{align}
and thus the cost to achieve $\epsilon$ is
\begin{align}
    \chi_l
    =\sum_{k=0}^l\binom{K+k-1}{k}
    \leq (l+1)\binom{K+l-1}{l}
    \leq (l+1)\frac{(K+l-1)^l}{l!}
    =O(l^K)
    =O\left(\left(\log\frac{1}{\epsilon}\right)^{K/2}\right).
\end{align}
Therefore, the upper bound of the cost from the concentration inequality increases exponentially in $K$.

\end{widetext}

\section{Bayesian test}\label{app:bayesian}
\begin{figure}[t]
\includegraphics[width=240px]{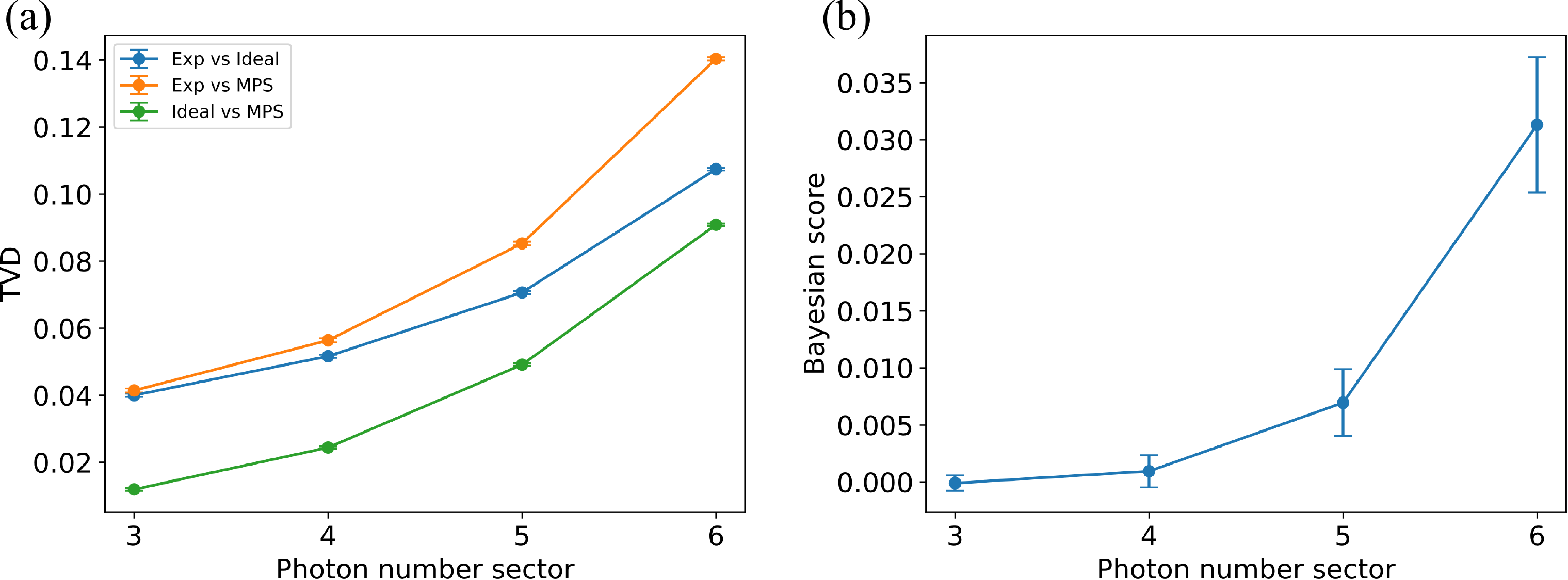}
\caption{(a) TVD between the experimental sampler, MPS sampler ($\chi=20$), and the ideal sampler. (b) The Bayesian score of the MPS sampler.}
\label{fig:bayesian}
\end{figure}

As mentioned in the main text, Refs.~\cite{zhong2021phase, madsen2022quantum, deng2023gaussian} employ the Bayesian test as a benchmark.
In this appendix we show that the test is not suitable for our purpose and our sampler because our sampler is closer to the ground-truth distribution than the experimental sampler is but our sampler is far from the experimental sampler.
To see this, we compute the TVD of all the pairs between the ideal, MPS, and experimental samplers, as illustrated in Fig.~\ref{fig:bayesian}(a).
The figure clearly shows that while the MPS sampler is a good approximation of the ground-truth distribution, it still has a large distance to the experimental sampler.
Therefore, as shown in the figure, even though our sampler approximates the ideal distribution better than the experiments in TVD, it fails to pass the Bayesian test, where the Bayesian score is defined as
\begin{align}
    \text{score}\equiv \frac{1}{N_s}\sum_{j=1}^{N_s}\log \frac{\Pr^{(G)}(\bm{m}_j)\Pr^{(s)}(N)}{\Pr^{(s)}(\bm{m}_j)\Pr^{(G)}(N)},
\end{align}
where $G$ stands for the ground truth and $\Pr^{(i)}(\bm{m})$ is the output probability of obtaining an outcome $\bm{m}$ a sampler $i$, which is either ideal or a mock-up sampler, and $\Pr^{(i)}(N)$ is the probability of obtaining total $N$ photons.
Thus, a positive score means that the ground-truth distribution is a better approximation than the mock-up distribution.
Consequently, the failure of our sampler to pass the Bayesian test does not imply that our sampler is not a good approximation of the ground-truth distribution.

\section{Intermediate-scale experiments}\label{app:intermediate}
\begin{figure}[t]
\includegraphics[width=240px]{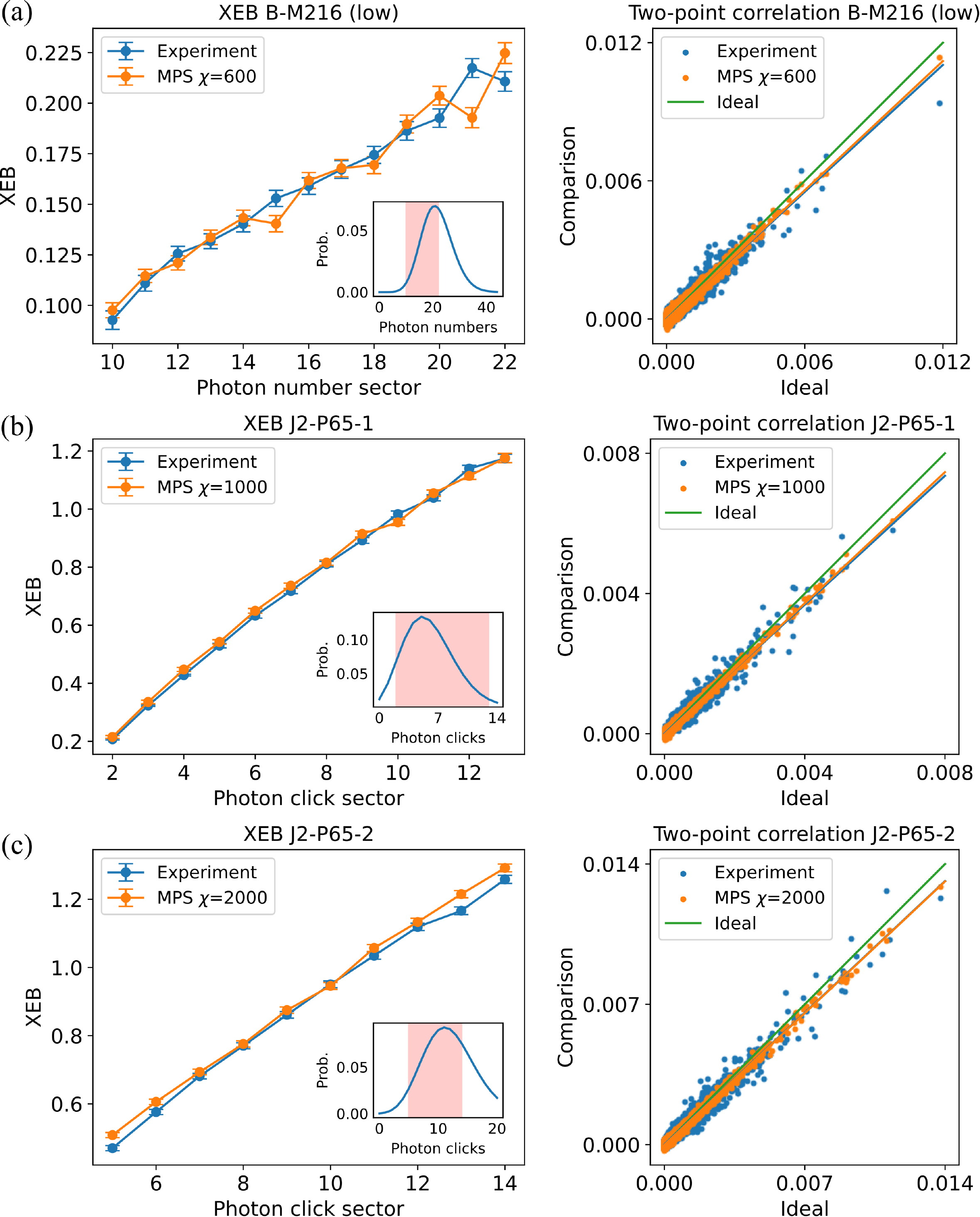} 
\caption{XEB and two-point correlation function for (a) Borealis $M=216$ (low), (b) Jiuzhang2.0's P65-1 with $M=144$, and (c) Jiuzhang2.0's P65-2 with $M=144$.}
\label{fig:intermediate}
\end{figure}

In this appendix we display an additional plot for intermediate cases:
Borealis's $M=216$ (low) case and Jiuzhang2.0's $M=144$ with P65-1 and P65-2.
For each case we choose $\chi=600, 1000, 2000$ with $d=4$ ($d=6$ is used for sampling by patching the tensors.), respectively. We  numerically demonstrate that the bond dimensions are sufficient to achieve better scores than experiments in XEB and two-point correlation benchmarking, as exhibited in Fig.~\ref{fig:intermediate}.
Also, each case entails an error of $0.0114, 0.0078, 0.0162$, respectively.
For all cases with the chosen bond dimensions, our MPS simulator achieves comparable XEB scores and two-point correlation functions.

\section{Implementation}\label{app:implementation}

We implement our simulation algorithm using Python. Specifically, GPU computations are optimized by using the CuPy library, and distributed computation is achieved by using the Message Passing Interface (MPI) using the MPI for Python library. We also develop a custom CUDA kernel for a minor subroutine in CUDA C++, which interfaces with Python through CuPy. Part of the algorithm is taken or modified from the source code of the Strawberry Fields library~\cite{killoran2019strawberry} as well as The Walrus library~\cite{gupt2019walrus}.

\begin{figure*}[t]
\includegraphics[width=510px]{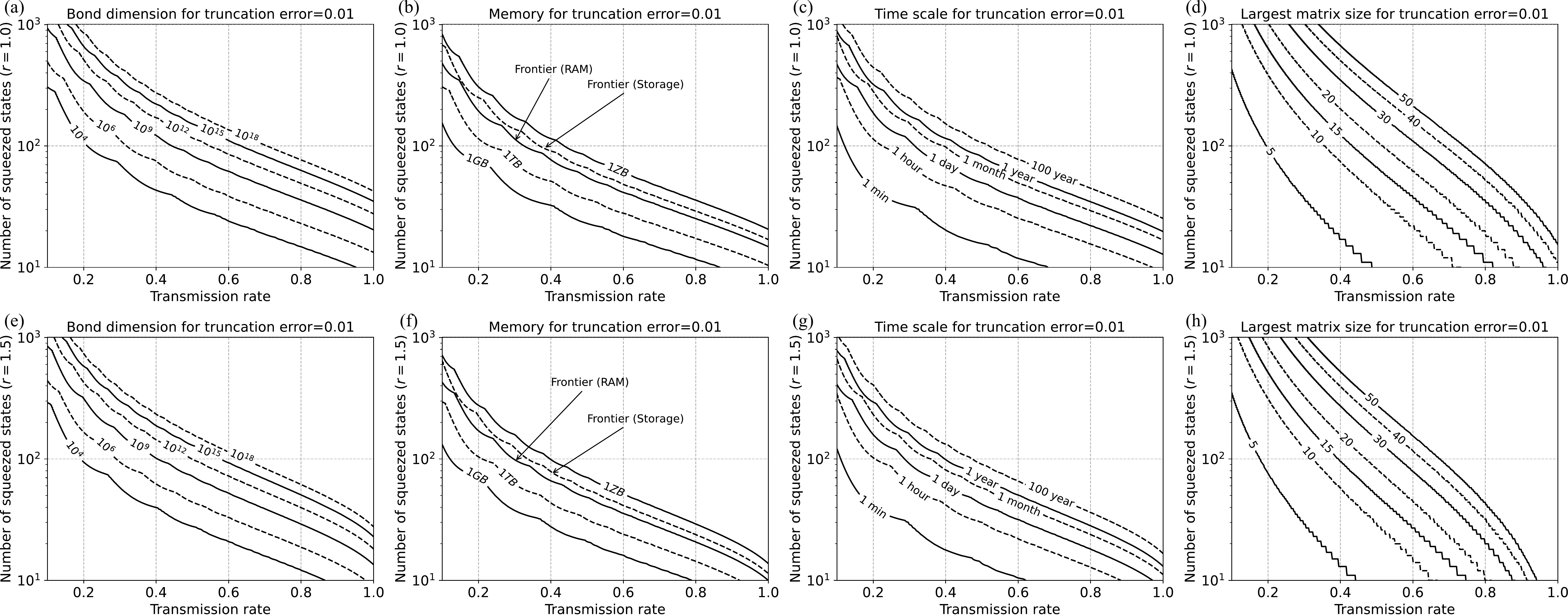}
\caption{Computational cost estimate for constructing an MPS that achieves truncation error $0.01$ with squeezing parameter (a-d) $r=1.0$ and (e-h) $r=1.5$.
The costs are characterized by (a)(e) MPS bond dimension, (b)(f) memory usage, (c)(g) time scale, and (d)(h) the matrix size for which we have to compute the hafnian.}
\label{fig:estimate_app}
\end{figure*}

The simulation algorithm uses a single GPU for the computation and storage of a single-mode MPS tensor. During the MPS calculation, all tensors are computed independently on different GPUs, which fills tensor entries with appropriately calculated hafnian values. Hafnians of equal-sized square matrices are computed in parallel for numerical efficiency, and this is possible because the hafnian calculation algorithm is data-independent. To avoid impractical memory costs, we limit the maximum number of parallel hafnians to a value dependent on the size of the input matrices, and we loop over subbatches to complete all hafnians. After a tensor is completed, the tensor is saved to the local SSD for later use during sampling.

Additionally, since the computational cost for different modes varies dramatically, GPUs that completed the designated tensor calculations are used to accelerate the computation of more costly tensors. Specifically, when challenging tensors have too many parallel hafnians to compute and other GPUs are available, subbatches are sent via the communication fabric for computation, and the results are later collected.

For the sampling algorithm, the computed single-mode MPS tensors are loaded to each GPU. During the sampling procedure, a vector is passed from one mode to the next via the communication fabric after some operations with local tensors for sampling. After the vector is sent to the next GPU, a new tensor is received from the previous GPU for new samples, resulting in a stream of samples that propagate through the chain. This process is also performed in a parallel manner via batch parallel random displacement generation, matrix multiplications, and weighted random choices. For the largest simulations, 100 samples are processed in parallel on a single GPU for numerical efficiency and reasonable memory costs. Therefore, $100 \times M$ samples are processed in parallel on $M$ GPUs.

\begin{figure*}[t]
\includegraphics[width=500px]{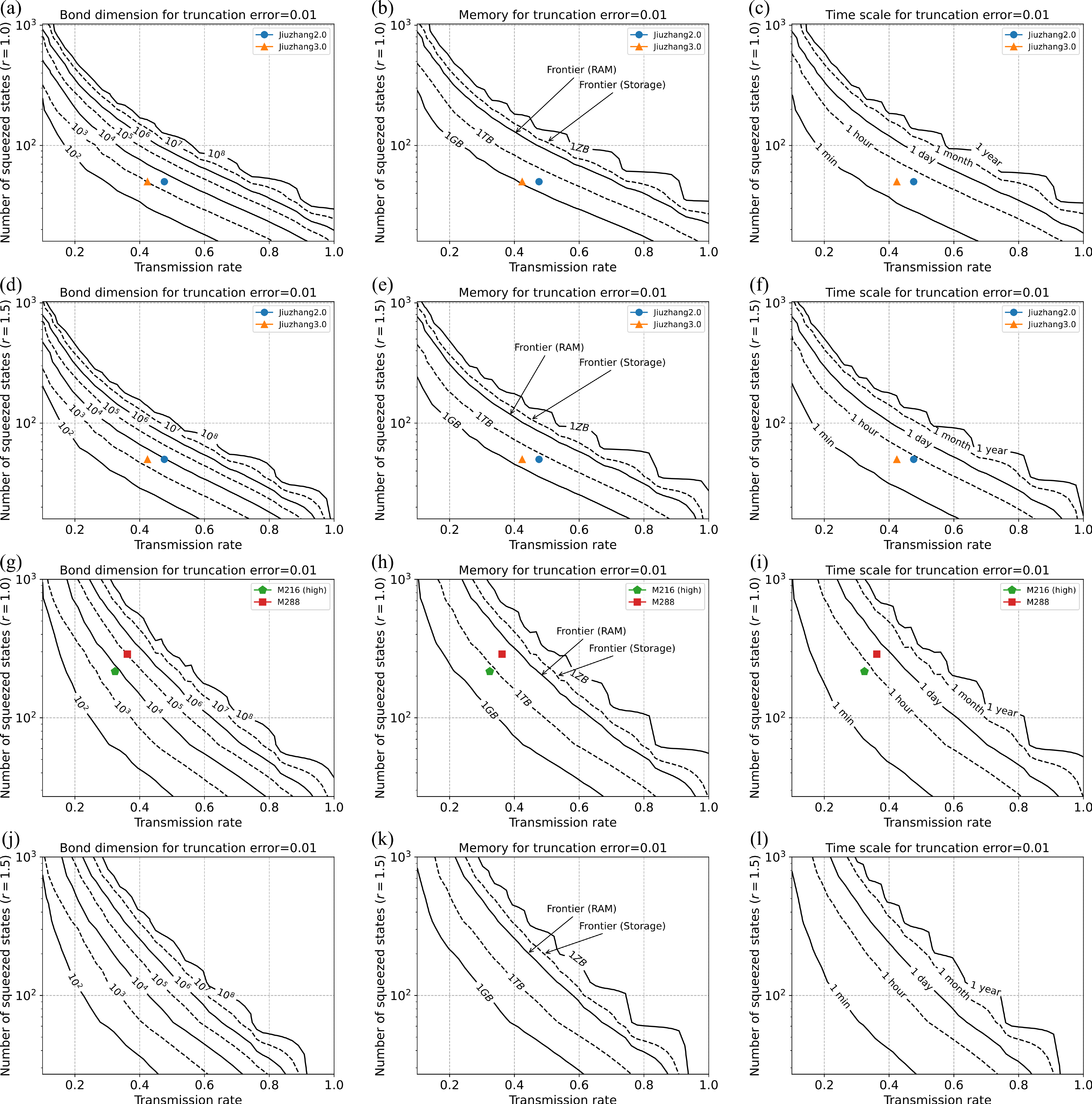}
\caption{Computational cost estimate for constructing an MPS that achieves truncation error $0.01$ with squeezing parameter (a-c) and (g-i) $r=1.0$ and (d-f) and (j-l) $r=1.5$.
We consider the USTC's setup in (a-f) and Xanadu's setup in (g-l).
The costs are characterized by (a)(d)(g)(j) MPS bond dimension, (b)(e)(h)(k) memory usage, and (c)(f)(i)(l) time scale.
The dots represent the relevant parameters in each experiment, and we do not present them in (j-l) because Xanadu's largest squeezing parameter is lower than $r=1.5$.
We averaged over ten different random circuits.}
\label{fig:estimate_both}
\end{figure*}

\section{Computational cost estimation}\label{app:estimate}
In this appendix, we further provide estimates of the computational cost of our algorithm for simulating larger systems and that for simulating systems that take into account the experimental architecture.
First, as in the main text, to make the analysis simple and consistent, we again assume the same conditions for the setup: (i)~input squeezing parameters and loss rates are uniform across the modes, (ii)~all the inputs are squeezed states without any vacuum states although they are not necessarily satisfied in the current experiments.
We emphasize that while the USTC's experiments used a different number of squeezed state inputs, 50, from the number of modes, 144, i.e., many of the inputs were the vacuum, in our analysis, we set them to be equal to remove the dependence on the position of input squeezed states.

We again consider three different circuit ensembles: (a) the worst-case circuits as in Fig.~\ref{fig:bipartition}, (b) the circuits similar to the USTC's experiments, and (c) the circuits similar to Xanadu's experiments.
More specifically, for (b) and (c), we model the USTC's experiment using the same structure with local beam splitters and assume that local beam splitters follow Haar-random unitaries.
We also model Xanadu's experiment using the same 3D structure with local Haar-random random unitary beam splitters.

We present the estimate for the worst-case circuit in Fig.~\ref{fig:estimate_app} and the one for the circuit modeling the experiments in Fig.~\ref{fig:estimate_both}.
Here, the required bond dimension for achieving a certain truncation error is obtained by considering the bipartition in Fig.~\ref{fig:bipartition}(c) and using Eq.~\eqref{eq:truncation} for the worst-case circuits analytically and directly numerically computing all the largest singular values for the USTC and Xanadu's setup.
The required memory usage is estimated by 
\begin{align}
    &(8~\text{bytes}) \times (\text{bond dimension})^2\times (\text{the number of modes}) \nonumber \\
    &\times (\text{local Hilbert space dimension}),
\end{align}
where $8$ bytes is for Complex64 variables, and we set local Hilbert space dimension $4$ as our simulation.
The time scale represents the time cost to prepare the MPS by counting the number of matrix elements without taking into account the complexity of computing each element (see below), which is estimated by
\begin{align}
    &(600~\text{secs})\times \left(\frac{\text{bond dimension}}{10000}\right)^2 \nonumber \\
    &\times (\text{local Hilbert space dimension}),
\end{align}
where $600$ secs is to use our simulation time as a benchmark, which used $\chi=10^4$ bond dimension.
We note that the time scale can vary for different computing devices or be significantly reduced by employing more resources.
Finally, we obtain the largest matrix size for which we have to compute the hafnian to construct an MPS, which is equal to $2l$ in Eq.~\eqref{eq:truncation}.
This corresponds to the cost of computing each element of MPS that was not taken into account in the time scale.
In fact, as presented in the figure, for the relevant regimes for the current experiments, the matrix size is not too large to compute in a reasonable time.
Also, Ref.~\cite{bjorklund2019faster} shows that computing the hafnian of matrix size less than $50\times 50$ is still possible with a reasonable computational cost.
Note that although we do not present the matrix size estimate in Fig.~\ref{fig:estimate_both} because of numerical uncertainty, based on the worst-case circuit results, we expect that the matrix size is not an immediate obstacle for the current experimental parameters.

\begin{acknowledgements}
We thank Yu-Hao Deng and Chao-Yang Lu for providing the dataset of the Jiuzhang3.0 experiment and for interesting and fruitful discussions. 
We thank Nicol\'{a}s Quesada and Jake Bulmer for helpful discussions.

This research used the resources of the Argonne Leadership Computing Facility, which is a U.S. Department of Energy (DOE) Office of Science User Facility supported under Contract DE-AC02-06CH11357. The authors are also grateful for the support of the University of Chicago Research Computing Center for assistance with the numerical simulations reported in this paper.
We acknowledge The Walrus Python library for the open source of Gaussian boson sampling algorithms~\cite{gupt2019walrus} and the Strawberry Fields library~\cite{killoran2019strawberry}.

M.L. acknowledges support from DOE Q-NEXT. 
Y.A. acknowledges support from the U.S. Department of Energy Office of Science under contract DE-AC02-06CH11357 at Argonne National Laboratory and Defense Advanced Research Projects Agency (DARPA) under Contract No. HR001120C0068. 
B.F acknowledges support from AFOSR (FA9550-21-1-0008).  This material
is based upon work partially supported by the National Science
Foundation under Grant CCF-2044923 (CAREER) and by the U.S. Department
of Energy, Office of Science, National Quantum Information Science
Research Centers (Q-NEXT), as well as by DOE QuantISED grant
DE-SC0020360. 
L.J. acknowledges support from the ARO(W911NF-23-1-0077), ARO MURI (W911NF-21-1-0325), AFOSR MURI (FA9550-19-1-0399, FA9550-21-1-0209, FA9550-23-1-0338), NSF (OMA-1936118, ERC-1941583, OMA-2137642, OSI-2326767, CCF-2312755), NTT Research, Packard Foundation (2020-71479).
\end{acknowledgements}

\bibliography{reference.bib}

\end{document}